\documentclass{lmcs} %%% last changed 2014-08-20
\pdfoutput=1

\usepackage{lastpage}
\renewcommand{\dOi}{14(4:20)2018}
\lmcsheading{}{1--\pageref{LastPage}}{}{}%
{Jan.~31,~2017}{Dec.~07,~2018}{}

\keywords{One-counter automata, disequality tests, reachability, Freeze LTL, Presburger arithmetic}

\ACMCCS{[{\bf Theory of computation}]: Formal languages and automata theory---Automata over infinite objects}

\usepackage{hyperref}
\usepackage{amsfonts}
\usepackage{amsmath}
\usepackage{amssymb}
\usepackage{amsthm}
\usepackage{microtype} %if unwanted, comment out or use option "draft"
\usepackage{latexsym}
\usepackage{stmaryrd}
\usepackage{algorithmic}
\usepackage{mathtools}
\usepackage{tikz}
\usetikzlibrary{automata,positioning}
\usepackage{pgflibraryshapes}

\newcommand{\vx}{\boldsymbol{x}}
\newcommand{\vk}{\boldsymbol{k}}
\newcommand{\N}{\mathbb{N}}
\newcommand{\Z}{\mathbb{Z}}
\newcommand{\cC}{\mathcal{C}}

\newcommand{\xtransrightarrow}[1]{\mathrel{%
  \vphantom{\xrightarrow{#1}}%
  \smash{\xrightarrow{#1}}%
  \vphantom{\to}^*}%
}
\newcommand{\FLTL}{\text{LTL}^\downarrow}
\newcommand{\oX}{\mathop{\mathsf{X}}}
\newcommand{\oU}{\mathrel{\mathsf{U}}}

\newcommand{\pathstart}[1]{\ensuremath{\operatorname{start}(#1)}}
\newcommand{\pathend}[1]{\ensuremath{\operatorname{end}(#1)}}
\newcommand{\pathweight}[1]{\ensuremath{\operatorname{weight}(#1)}}

\newcommand{\pathlen}[1]{\ensuremath{|#1|}}

\newcommand{\simplecycleall}{\ensuremath{\operatorname{SC}}}
\newcommand{\simpcycleposall}{\ensuremath{\operatorname{SC}^{+}}}
\newcommand{\simpcyclenegall}{\ensuremath{\operatorname{SC}^{-}}}

\newcommand{\cyclemark}[1]{\ensuremath{\underline{#1}}}

\newcommand{\pathunfold}[1]{\ensuremath{\operatorname{unfold}(#1)}}

\begin{document}

%%\title[Instructions]{Instructions for Authors\\How to prepare papers
%%  for LMCS using \MakeLowercase{\texttt{lmcs.cls}}\rsuper*\\Version of 
%%  2014-02-01}
%%\titlecomment{{\lsuper*}OPTIONAL comment concerning the title, \eg, 
%%  if a variant or an extended abstract of the paper has appeared elsewhere.}
\title{Model Checking Flat Freeze LTL on One-Counter Automata}

\author[A.~Lechner]{Antonia Lechner\rsuper{1}}	%required
\address{\lsuper{1}Computer Science Department, Universit\'e Libre de Bruxelles, Belgium}
%in case full address should be required
%\address{Computer Science Department, CP 212, ULB Campus de la Plaine, Boulevard du Triomphe, 1050 Bruxelles, Belgium}	%required
\email{antonia.lechner@ulb.ac.be}  %optional
%\thanks{thanks 1, optional.}	%optional

\author[R.~Mayr]{Richard Mayr\rsuper{2}}	%optional
\address{\lsuper{2}School of Informatics, LFCS, University of Edinburgh, UK}
%in case full address should be required
%\address{School of Informatics, LFCS, University of Edinburgh, 10 Crichton Street, Edinburgh, EH8 9AB, UK}	%optional
\urladdr{http://homepages.inf.ed.ac.uk/rmayr/}
%\thanks{thanks 2, optional.}	%optional

\author[J.~Ouaknine]{Jo\"el Ouaknine\rsuper{3}}	%optional
\address{\lsuper{3}Max Planck Institute for Software Systems, Saarbr\"ucken, Germany}
%in case full address should be required
%\address{MPI-SWS, Campus E1 4, D-66123 Saarbr\"ucken, Germany}	%optional
\email{joel@mpi-sws.org}  %optional
%\thanks{thanks 3, optional.}	%optional

\author[A.~Pouly]{Amaury Pouly\rsuper{3}}	%optional
\address{\vskip-7pt} % Max Planck Institute for Software Systems, Saarbr\"ucken, Germany}
%in case full address should be required
%\address{MPI-SWS, Campus E1 4, D-66123 Saarbr\"ucken, Germany}	%optional
\email{pamaury@mpi-sws.org}  %optional
%\thanks{thanks 3, optional.}	%optional

\author[J.~Worrell]{James Worrell\rsuper{4}}	%optional
\address{\lsuper{4}Department of Computer Science, University of Oxford, UK}
%in case full address should be required
%\address{Department of Computer Science, University of Oxford, Wolfson Building, Parks Road, Oxford, OX1 3QD, UK}	%optional
\email{james.worrell@cs.ox.ac.uk}
%\urladdr{name3@url3\quad\rm{(optionally, a web-page can be specified)}}  %optional
%\thanks{thanks 3, optional.}	%optional

%% etc.

%% required for running head on odd and even pages, use suitable
%% abbreviations in case of long titles and many authors:

%%%%%%%%%%%%%%%%%%%%%%%%%%%%%%%%%%%%%%%%%%%%%%%%%%%%%%%%%%%%%%%%%%%%%%%%%%%

%% the abstract has to PRECEDE the command \maketitle:
%% be sure not to issue the \maketitle command twice!

%%\begin{abstract}
  \begin{abstract}
  Freeze LTL is a temporal logic with registers that is suitable for
  specifying properties of data words.  In this paper we study the
  model checking problem for Freeze LTL on one-counter automata. This
  problem is known to be undecidable in general and
  PSPACE-complete for the special case of deterministic one-counter
  automata. Several years ago, Demri and Sangnier investigated the
  model checking problem for the flat fragment of Freeze LTL on
  several classes of counter automata and posed the decidability of
  model checking flat Freeze LTL on one-counter automata as an open
  problem. In this paper we resolve this problem positively, utilising
  a known reduction to a reachability problem on one-counter automata
  with parameterised equality and disequality tests. Our main technical
  contribution is to show decidability of the latter problem by
  translation to Presburger arithmetic.
\end{abstract}

%%% Local Variables:
%%% mode: latex
%%% TeX-master: "Main-Revised.tex"
%%% End:

%%\end{abstract}

\maketitle

%% start the paper here:

\section{Introduction}
Runs of infinite-state machines, such as counter automata, can
naturally be seen as \emph{data words}, that is, sequences in which
each position is labelled by a letter from a finite alphabet and a
datum from an infinite domain.  Freeze LTL is an extension of Linear
Temporal Logic with registers and a binding mechanism,
which has been introduced to specify properties of data
words~\cite{DemriL09,DemriLN07,French03,LisitsaP05}.  The registers
allow to compare data at different positions along the same
computation.

An example of a freeze LTL formula is
\begin{gather} \mathop{\mathsf{F}}(v \wedge {\downarrow_r \mathsf{XF}(v \wedge {\uparrow_r})}) \, . 
\label{eq:flat}
\end{gather}
Evaluated on a run of a one-counter automaton, this formula is true if
and only if there are at least two different positions in the run
which both have control state $v$ and the same counter value.
Intuitively the operator $\downarrow_r$ binds the current counter
value to register $r$, while the operator $\uparrow_r$ tests whether
the current counter value is equal to the content of register $r$.

This paper concerns the model checking problem for Freeze LTL on
one-counter automata.  It is known that this problem is undecidable in
general and PSPACE-complete if one restricts to \emph{deterministic}
one-counter automata~\cite{DemriLS10}.  Rather than restricting the
class of one-counter automata, one can seek to identify decidable
syntactic fragments of Freeze LTL.  This approach was pursued
in~\cite{DemriS10}, which studied the \emph{flat} fragment of Freeze
LTL.  The flatness condition places restrictions on the occurrence of
the binding construct $\downarrow_r$ in relation to the until operator
(see Section~\ref{sec:flat} for details).  For example, in a flat
formula in negation normal form the binding operator $\downarrow_r$
can occur within the scope of $\mathsf{F}$ but not $\mathsf{G}$.
(Thus formula (\ref{eq:flat}) is flat.)  The flatness restriction for
Freeze LTL has a similar flavour to the respective flatness
restrictions for constraint LTL~\cite{ComonC00} and for Metric
Temporal Logic~\cite{BouyerMOW08}.

Demri and Sangnier~\cite{DemriS10} considered the decidability of
model checking flat Freeze LTL across a range of different
counter-machine models.  For one-counter automata they
showed decidability of model checking for a certain fragment of flat
Freeze LTL and they left open the problem of model checking flat
Freeze LTL in general.  

The approach taken in~\cite{DemriS10} was to reduce the model checking
problem for fragments of Freeze LTL on counter automata to
reachability problems in counter automata augmented with certain kinds
of parameterised tests.  Specifically they reduce the model checking
problem for flat Freeze LTL on one-counter automata to the problem of
deciding reachability of Buchi objectives on one-counter automata
extended with parameterised equality and disequality tests.  The
latter problem considers one-counter automata whose transitions may be
guarded by equality or disequality tests that compare the counter
value to integer-valued parameters, and it asks whether there exist
parameter values such that there is an infinite computation that
visits an accepting location infinitely many times.  The parameterised
tests are used to handle register binding in freeze LTL.  The main
technical contribution of this paper is to show decidability of the
latter reachability problem by reduction to the decision problem for
Presburger arithmetic.  We thereby show that the model checking
problem for flat Freeze LTL on one-counter automata is decidable.

A related work is~\cite{HKOW}, which considers one-counter automata
with parameterised updates and equality tests.  It is shown
in~\cite{HKOW} that reachability in this model is inter-reducible with
the satisfiability problem for quantifier-free Presburger arithmetic
with divisibility, and therefore decidable.  In contrast
to~\cite{HKOW}, in the present paper the counter automata do not have
parameterised updates but they do have parameterised disequality tests.  The
results in this paper do not appear to be straightforwardly reducible
to those of~\cite{HKOW} nor \emph{vice versa}.  Both reachability
problems can be seen as special cases of a long-standing open
problem identified by Ibarra \emph{et al.}~\cite{IJTW93-icalp}, which
asks to decide reachability on a class of automata with a single
integer-valued counter, sign tests, and parameterised updates.

%%% Local Variables:
%%% mode: latex
%%% TeX-master: "Main-Revised.tex"
%%% End:

\section{Preliminaries}

\subsection{One-Counter Automata with 
Equality and Disequality Tests}

We consider automata with a single counter that ranges over the
nonnegative integers, equipped with both equality and disequality
tests on counter values. Formally, a \emph{one-counter automaton}
(1-CA) is a tuple $\mathcal{C}=(V,E,\lambda,\tau)$, where $V$ is a
finite set of \emph{states}, $E \subseteq V \times V$ is a finite set
of \emph{edges} between states, $\lambda : E \rightarrow Op$ labels
each edge with an element from
$Op = \{ \mathrm{add}(a): a \in \Z\} \cup \{ \mathrm{eq}(a) : a \in \N
\}$, and $\tau : V \rightarrow 2^{\mathbb{N}}$ maps each state $v$ to
a finite set $\tau(v)$ of \emph{invalid counter values} at state $v$.
Intuitively the operation $\mathrm{add}(a)$ adds $a$ to the counter
and $\mathrm{eq}(a)$ tests the counter for equality with $a$.  The
association of invalid counter values with each state can be seen as a
type of disequality test.  This last feature is not present in
classical presentations of 1-CA, but we include it here to facilitate
our treatment of Freeze LTL.

For any edge $e=(v,v')$, define $\pathstart{e}=v$ and
$\pathend{e}=v'$; moreover write $\pathweight{e}=a$ if
$\lambda(e)=\mathrm{add}(a)$ and $\pathweight{e}=0$ if
$\lambda(e)=\mathrm{eq}(a)$. A \emph{path} $\gamma$ is a finite word
on the alphabet $E$: $\gamma=e_1\ldots e_n$ such that
$\pathend{e_i}=\pathstart{e_{i+1}}$ for all $1 \leqslant i < n$. The
\emph{length} of $\gamma$, denoted $\pathlen{\gamma}$, is $n$. The
\emph{state sequence} of $\gamma$ is $\pathstart{e_1}$,
$\pathend{e_1}$, $\pathend{e_2}$, \ldots, $\pathend{e_n}$. The
\emph{start} of $\gamma$, denoted $\pathstart{\gamma}$, is
$\pathstart{e_1}$. The \emph{end} of $\gamma$, denoted
$\pathend{\gamma}$, is $\pathend{e_n}$.  A path is \emph{simple} if it
contains no repeated states.  The \emph{weight} of $\gamma$, denoted
by $\pathweight{\gamma}$, is $\sum_{i=1}^n\pathweight{e_i}$. A
\emph{subpath} $\gamma'$ of $\gamma$ is any factor of $\gamma$:
$\gamma'=e_ie_{i+1}\ldots e_j$. If $\gamma$ and $\gamma'$ are two
paths such that $\pathend{\gamma}=\pathstart{\gamma'}$,
$\gamma\gamma'$ is the concatenation of both paths.

A \emph{cycle} $\omega$ is a path such that
$\pathstart{\omega}=\pathend{\omega}$.  A cycle is \emph{simple} if
it has no repeated states except for the starting point, which
appears twice.  A cycle is \emph{positive} if it has positive weight,
\emph{negative} if it has negative weight and \emph{zero-weight} if it
has weight zero.  We denote by
$\omega^k=\underbrace{\omega\omega\ldots\omega}_{k\text{ times}}$ the
sequence of $k$ iterations of the cycle $\omega$.

A \emph{configuration} of a 1-CA $\mathcal{C}=(V,E,\lambda,\tau)$ is a pair $(v,c)$ with $v \in V$ and $c \in \Z$.
Intuitively, $(v,c)$ corresponds to the situation where the 1-CA is in state $v$ with counter value $c$. 
Configurations $(v,c)$ with $c \geqslant 0$ and $c\not\in\tau(v)$
are called \emph{valid}, otherwise they are said to be \emph{invalid}.
The edge relation $E$ induces an unlabelled transition relation between
configurations:  for any two configurations $(v,c)$ and
$(v^\prime,c^\prime)$, there is a transition
$(v,c) \longrightarrow (v^\prime,c^\prime)$ if and only if there is an
edge $e \in E$ such that $\pathstart{e} = v$,
$\pathend{e} = v^\prime$, and $\pathweight{e} = c^\prime -c$.  We will
sometimes write $(v,c) \xrightarrow{\text{ }e\text{ }} (v^\prime,c^\prime)$ for such a
transition.  The transition is \emph{valid} if both $(v,c)$ and $(v',c')$ are valid configurations 
and $c = a$ if $\lambda(e) = \mathrm{eq}(a)$. Otherwise such a transition is
\emph{invalid}.

A \emph{computation} $\pi$ is a (finite or infinite) sequence of transitions:
  \[\pi=(v_1,c_1)\longrightarrow(v_2,c_2)\longrightarrow (v_3,c_3) \longrightarrow \cdots \]
  We write
  $\lvert \pi \rvert$ for the length of $\pi$. If $(v_1,c_1) \xrightarrow{\text{ }e_1\text{ }} (v_2,c_2) \xrightarrow{\text{ }e_2\text{ }} \cdots \xrightarrow{e_{n-1}} (v_n,c_n)$ is a finite computation, we will also write it as $(v_1,c_1) \xtransrightarrow{\text{ }\gamma\text{ }} (v_n,c_n)$, where $\gamma = e_1 e_2 \ldots e_{n-1}$, or simply $(v_1,c_1) \longrightarrow^* (v_n,c_n)$.
A computation $\pi$ is \emph{valid} if all transitions in the sequence are valid, otherwise it is \emph{invalid}.
\textcolor{black}{If $\pi$ is invalid, an \emph{obstruction} is a configuration $(v_i,c_i)$ such that either $(v_i,c_i)$ is invalid or
$(v_i,c_i)$ is not the final configuration in $\pi$ and
$(v_i,c_i) \longrightarrow (v_{i+1},c_{i+1})$ is an invalid transition.}

Given a path $\gamma$ and a counter value $c\in\Z$, the \emph{path computation} $\gamma(c)$ is
the (finite) computation starting at $(\pathstart{\gamma},c)$ and following the sequence of transitions that correspond to the edges in $\gamma$.

A \emph{one-counter automaton with parameterised tests} is a tuple
$(V,E,X,\lambda,\tau)$, where $V$, $E$ and $\lambda$ are defined as
before for 1-CA, $X$ is a set of nonnegative integer parameters,
$Op = \{ \mathrm{add}(a) : a \in \Z\} \cup
\{\mathrm{eq}(a),\mathrm{eq}(x) : a \in \N, x \in X\}$ includes
parameterised equality tests (but not parameterised updates), and
$\tau : V \rightarrow 2^{\N \cup X}$ includes parameterised
disequality tests. Note that $\tau(v)$ is still required to be finite
for each $v \in V$.

For a given 1-CA $\mathcal{C} = (V,E,\lambda,\tau)$, an initial configuration $(v,c)$ and a target configuration $(v^\prime,c^\prime)$, the \emph{reachability} problem asks if there is a valid computation from $(v,c)$ to $(v^\prime,c^\prime)$. When $\cC$ has sets $F_1,\ldots,F_n \subseteq V$ of final states and an initial configuration $(v,c)$, the \emph{generalised repeated control-state reachability} problem asks if there is a valid infinite computation from $(v,c)$ which visits at least one state in each $F_i$ infinitely often. 

For a 1-CA $\cC=(V,E,X,\lambda,\tau)$ with parameterised tests,
initial configuration $(v,c)$, and target configuration
$(v^\prime,c^\prime)$, the \emph{reachability} problem asks if there
exist values for the parameters such that there is a computation from
$(v,c)$ to $(v^\prime,c^\prime)$. Similarly, in the case where $\cC$
has sets $F_1,\ldots,F_n \subseteq V$ of final states and an initial
configuration $(v,c)$, the \emph{generalised repeated control-state
  reachability} problem asks if there exist values for the parameters
such that substituting these values satisfies the generalised repeated
control-state reachability condition above.

%%%additions from thesis:

Note that in our model of 1-CA, equality tests are defined on
transitions (via the function $\lambda$) while disequality tests are
defined on states (via the function $\tau$).  While this asymmetry may
seem unnatural, it is technically convenient for the subsequent
proofs.  By contrast in the model of 1-CA in~\cite{DemriS10} both
equality and disequality tests are defined on transitions.  This model
also allows multiple edges between states, which is excluded in our
formalism.  Nevertheless,
from the point of view of reachability and repeated reachability the
model of 1-CA that we use and that of~\cite{DemriS10} are easily seen
to be equivalent (so that an algorithm for one type of 1-CA will work
for the other type with only a polynomial overhead).  For example,
\textcolor{black}{compare the 1-CA in Figure~\ref{fig:1ca-with-tau} (in figures, we
write $+a$ for $\mathrm{add}(a)$ and $=a?$ for $\mathrm{eq}(a)$) and
the 1-CA in Figure~\ref{fig:1ca-without-tau}, which has disequality
tests defined on transitions rather than states, as well as multiple
edges between $u_5$ and $u_6$.  Then the computation
\[ (v_1,0) \longrightarrow (v_2,10) \longrightarrow (v_2,8) \longrightarrow (v_3,5) \longrightarrow (v_5,1) 
\longrightarrow (v_6,1) \]
for the 1-CA in Figure~\ref{fig:1ca-with-tau} corresponds to the computation
\begin{align*}
  & (u_1,0) \longrightarrow (u_2,10) \longrightarrow (u_3,10) \longrightarrow (u_4,10) \longrightarrow (u_2,8) \\
  &  \longrightarrow (u_3,8) \longrightarrow (u_4,8) \longrightarrow (u_5,5) \longrightarrow (u_6,1) \longrightarrow (u_7,1) 
\end{align*}
for the 1-CA in Figure~\ref{fig:1ca-without-tau}.}

%Note that while we have defined equality tests ($\lambda$) to be on
%transitions of a 1-CA, such that the counter is required to have a
%certain value for a transition with an equality test to be valid,
%disequality tests ($\tau$) are on the states of the 1-CA, such that
%the counter is required to not be contained in a certain set of values
%for any transition out of the current state to be valid. It may seem
%more intuitive to define both types of test on transitions instead,
%and in fact this is how 1-CA were defined in \cite{DemriS10}, where
%the above reachability problems for 1-CA with parameterised tests were
%stated as open. The same paper also allows multiple edges between any
%two states of a 1-CA. The reason why we chose to work with a model
%that has disequality tests on states and only at most one edge between
%any two given states is that it simplifies some of the following
%proofs.
%
%It is easy to see that the type of 1-CA in this chapter is equivalent
%to the one defined in \cite{DemriS10}, so that an algorithm for
%reachability for one of them also solves the same problem for the
%other. 
%%Clearly
%%$(v_1,0) \longrightarrow^* (v_6,1)$ if and only if
%%$(u_1,0) \longrightarrow^* (u_8,1)$.

\begin{figure}
  \centering
  \begin{tikzpicture}
    \node[state] (v1) {$v_1$};
    \node[state,label=below:{$\tau(v_2) = \{2,3\}$}] (v2) [right=2cm of v1] {$v_2$};
    \node[state] (v3) [right=2cm of v2] {$v_3$};
    \node[state] (v4) [above right=0.8cm and 1.2cm of v3] {$v_4$};
    \node[state] (v5) [below right=0.8cm and 1.2cm of v3] {$v_5$};
    \node[state] (v6) [below right=0.8cm and 1.2cm of v4] {$v_6$};
    \path[->]
    (v1) edge node [above] {$+10$} (v2)
    (v2) edge [loop above] node {$-2$} ()
         edge node [above] {$-3$} (v3)
    (v3) edge node [above left] {$-3$} (v4)
         edge node [below left] {$-4$} (v5)
    (v4) edge node [above right] {$=1?$} (v6)
    (v5) edge node [below right] {$=1?$} (v6);
  \end{tikzpicture}
  \caption{A simple 1-CA including a disequality test on the state $v_2$ and equality tests on the transitions $(v_4,v_6)$ and $(v_5,v_6)$.}
  \label{fig:1ca-with-tau}
\end{figure}
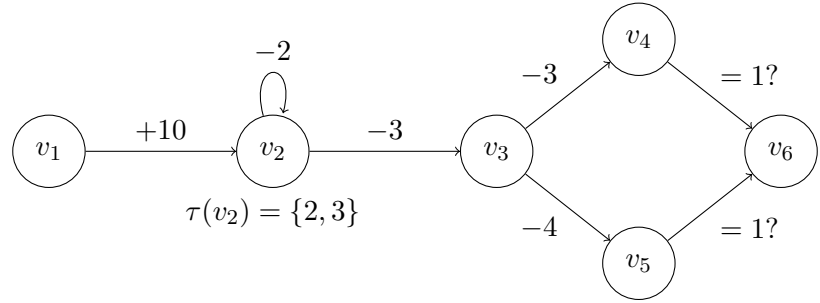

\begin{figure}
  \centering
  \begin{tikzpicture}
    \node[state] (u1) {$u_1$};
    \node[state] (u2) [right=1.5cm of u1] {$u_2$};
    \node[state] (u3) [above=1.3cm of u2] {$u_3$};
    \node[state] (u4) [right=1.5cm of u3] {$u_4$};
    \node[state] (u5) [right=1.5cm of u2] {$u_5$};
    \node[state] (u6) [right=1.5cm of u5] {$u_6$};
    \node[state] (u7) [right=1.5cm of u6] {$u_7$};
    \path[->]
    (u1) edge node [above] {$+10$} (u2)
    (u2) edge node [left] {$\neq 2?$} (u3)
%         edge node [above] {$\neq 2?$} (u5)
    (u3) edge node [above] {$\neq 3?$} (u4)
    (u4) edge node [above left] {$-2$} (u2)
         edge node [right] {$-3$} (u5)
%    (u5) edge node [above] {$\neq 3?$} (u6)
    (u5) edge [bend left] node [above] {$-3$} (u6)
         edge [bend right] node [below] {$-4$} (u6)
    (u6) edge node [above] {$=1?$} (u7);
  \end{tikzpicture}
  \caption{An automaton with disequality tests on transitions rather than states and with multiple edges which is equivalent to the one in Figure~\ref{fig:1ca-with-tau}.}
  \label{fig:1ca-without-tau}
\end{figure}
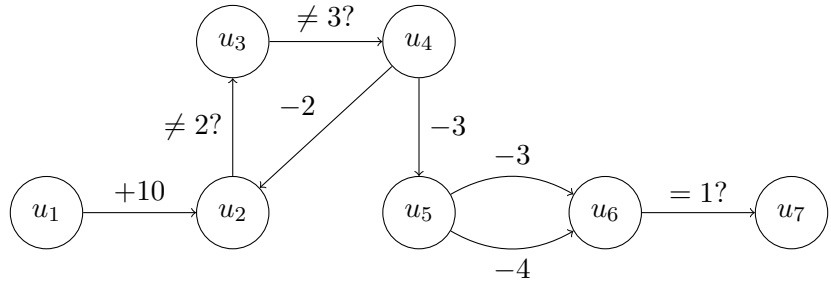

%As for multiple edges between states, consider the the 1-CA in Figure~\ref{fig:1ca-with-multiple-edges} which has

%\begin{figure}
%  \centering
% \begin{tikzpicture}
%   \node[state] (u1) {$u_1$};
% \node[state] (u2) [right=3cm of u1] {$u_2$};
%    \path[->]
%    (u1) edge [bend left] node [above right] {$+3$} (u2)
%         edge node [above] {$=5?$} (u2)
%         edge [bend right] node [below right] {$-2$} (u2);
%  \end{tikzpicture}
%  \caption{}
%  \label{fig:1ca-with-multiple-edges}
%\end{figure}

%\begin{figure}
%  \centering
%  \begin{tikzpicture}
%    \node[state] (v1) {$v_1$};
%    \node[state,style={scale=0.6}] (q1) [above right=1cm and 2cm of v1] {};
%    \node[state,style={scale=0.6}] (q2) [below=1cm of q1] {};
%    \node[state,style={scale=0.6}] (q3) [below right=1cm and 2cm of v1] {};
%    \node[state] (v2) [right=2cm of q2] {$v_2$};
%    \path[->]
%    (v1) edge node [above left] {$+3$} (q1)
%         edge node [above] {$=5?$} (q2)
%         edge node [below left] {$-2$} (q3)
%    (q1) edge node [above right] {$+0$} (v2)
%    (q2) edge node [above] {$+0$} (v2)
%    (q3) edge node [below right] {$+0$} (v2);
%  \end{tikzpicture}
%  \caption{}
%  \label{fig:1ca-without-multiple-edges}
%\end{figure}

%%% Local Variables:
%%% mode: latex
%%% TeX-master: "Main-Revised.tex"
%%% End:

\subsection{Model Checking Freeze LTL on One-Counter Automata}
\label{sec:flat}
\emph{Freeze LTL}~\cite{DemriLS10} is an extension of Linear Temporal
Logic that can be used to specify properties of data words. A data word is a (finite or infinite) sequence of symbols, each of which consists of a letter from a finite alphabet and another letter, often referred to as a \emph{datum}, from an infinite alphabet. Freeze
LTL is one of a variety of formalisms that arise by augmenting a temporal
or modal logic with variable binding.  Given a finite alphabet
$\Sigma$ and set of \emph{registers} $R$, the formulas of Freeze LTL
are given by the following grammar
\[ \varphi \quad ::= \quad a \quad | \quad \uparrow_r \quad | \quad
  \neg \varphi \quad | \quad \varphi \wedge \varphi \quad | \quad \oX
  \varphi \quad | \quad \varphi \oU \varphi 
\quad | \quad \downarrow_r \varphi \, , \] where
$a\in \Sigma$ and $r\in R$. In addition to the standard LTL
connectives, Freeze LTL contains an atomic freeze formula $\uparrow_r$
and a freeze operator $\downarrow_r$. We write $\FLTL$ for the set of
formulas of Freeze LTL.  A \emph{sentence} is a formula in which each
occurrence of a subformula $\uparrow_r$ is in the scope of an operator
$\downarrow_r$ (for the same register $r$).

In general, formulas of $\FLTL$ are interpreted over data words which
have $\Sigma$ as their finite alphabet and an arbitrary infinite
alphabet.  In this paper we are interested in a particular kind of
data word---namely those arising from valid computations of 1-CA.  We
directly define the semantics of $\FLTL$ over such computations,
assuming that the alphabet $\Sigma$ is the set of states of
the 1-CA and that the infinite alphabet for data words is $\N$.  In
this context $\downarrow_r$ can be seen as a binding construct that
stores in register $r$ the counter value at the current position in a
computation, while $\uparrow_r$ tests whether the counter value at the
current position is equal to the content of register $r$.  Formally,
define a \emph{register valuation} to be a partial function
$f:R\rightarrow\mathbb{N}$ and consider a valid infinite computation
\[ \pi = (v_1,c_1) \longrightarrow (v_2,c_2) \longrightarrow (v_3,c_3)
\longrightarrow \cdots \]
of a 1-CA $\mathcal{C}$.  We define a satisfaction
relation $\pi,i\vDash_f \varphi$ specifying when an $\FLTL$ formula
$\varphi$ is satisfied at position $i$ in $\pi$ under valuation $f$:
\begin{gather*}
  \begin{array}{rcl}
    \pi,i \vDash_f a & \xLeftrightarrow{\text{def}} & v_i = a \\
    \pi,i \vDash_f \, \, \uparrow_r & \xLeftrightarrow{\text{def}} & c_i = f(r) \\
    \pi,i \vDash_f \neg \varphi & \xLeftrightarrow{\text{def}} & \pi,i \not \vDash_f \varphi \\
    \pi,i \vDash_f \varphi_1 \vee \varphi_2 & \xLeftrightarrow{\text{def}} & \pi,i \vDash_f \varphi_1 \text{ or } \pi,i \vDash_f \varphi_2 \\
    \pi,i \vDash_f \oX \varphi & \xLeftrightarrow{\text{def}} & \pi,i+1 \vDash_f \varphi \\
    \pi,i \vDash_f \varphi_1 \oU \varphi_2 & \xLeftrightarrow{\text{def}} & 
\pi,j \vDash_f \varphi_2 \text{ for some } j \geqslant i \text{ and } \pi,k \vDash_f \varphi_1 
\text{ for all } i \leqslant k  < j\\
%    \pi,i \vDash_f \varphi_1 \oR \varphi_2 & \xLeftrightarrow{\text{def}} & \pi,j \vDash_f \varphi_2 \text{ for all } j \geqslant i, \\
%    & &\text{ or for some } j', \pi,j' \vDash_f \varphi_1 \text{ and for all } i \leqslant k \leqslant j', \pi,k \vDash \varphi_2 \\ 
    \pi,i \vDash_f \, \, \downarrow_r \varphi & \xLeftrightarrow{\text{def}} & \pi,i \vDash_{f[r \mapsto c_i]} \varphi 
  \end{array}
\end{gather*}
where $f[r \mapsto c]$ is the function that maps $r$ to $c$ and is
otherwise equal to $f$. Note that the clauses for the Boolean and LTL connectives are defined in the same way as for standard LTL.
%%%added these in the journal version
%We have omitted the clauses for the Boolean connectives.

An occurrence of a subformula in an $\FLTL$ formula is said to be
$\emph{positive}$ if it lies within the scope of an even number of
negations, otherwise it is \emph{negative}. The \emph{flat fragment}
of $\FLTL$ is the set of $\FLTL$ formulas such that in every positive
occurrence of a subformula $\varphi_1 \oU \varphi_2$ the binding
operator $\downarrow_r$ does not appear in $\varphi_1$, and in every
negative occurrence of such a subformula, $\downarrow_r$ does not
appear in $\varphi_2$ for any register $r$.

The negation of many natural $\FLTL$ specifications can be expressed
by flat formulas.  For example, consider the response property
$\mathop{\mathsf{G}}( \downarrow_r(\mathrm{req}\rightarrow
{\mathop{\mathsf{F}}(\mathrm{serve} \wedge {\uparrow_r})}))$,
expressing that every request is followed by a serve with the same
associated ticket.  \textcolor{black}{(Here $\mathsf{F}$ and $\mathsf{G}$ are the
``future'' and ``globally'' modalities, defined by
$\mathop{\mathsf{F}} \varphi := \mathbf{true} \oU \varphi$ and
$\mathop{\mathsf{G}} \varphi := \neg \mathop{\mathsf{F}} {\neg
  \varphi}$.)}  The negation of this formula is equivalent to
$\mathop{\mathsf{F}}( \downarrow_r(\mathrm{req}\wedge
{\mathop{\mathsf{G}}(\neg\mathrm{serve} \vee \neg{\uparrow_r})}))$.
The latter is easily seen to be flat after rewriting to the core
$\FLTL$ language with only the $\mathsf{U}$ temporal operator.

The main subject of this paper is the decidability of the following
model checking problem: given a 1-CA $\cC$, a valid configuration
$(v,c)$ of $\cC$, and a flat sentence $\varphi \in \FLTL$, does there
exist a valid infinite computation $\pi$ of $\cC$, starting at
$(v,c)$, such that $\pi,1 \vDash_\emptyset \varphi$?  Note that,
following~\cite{DemriS10}, we have given an existential formulation of
the model checking problem.  The model checking problem, as formulated
above, is equivalent to asking whether $\neg\varphi$ holds along all
valid infinite computations starting at $(v,c)$.

The model checking problem for flat $\FLTL$ on 1-CA was reduced to a
repeated reachability problem for 1-CA with parameterised tests
in~\cite[Theorem 15]{DemriS10}. The idea of the reduction is, given a
1-CA $\cC$ and a flat $\FLTL$ sentence $\varphi$ in negation normal
form, to construct a 1-CA with parameterised tests which is the
product of $\cC$ and $\varphi$. This product automaton includes a
parameter $x_r$ for each register $r$ that is mentioned in a
subformula of $\varphi$ of type $\downarrow_r \varphi'$. This is where
the restriction to the flat fragment of $\FLTL$ is crucial, since it
allows us to assume that the value stored in a register is never
overwritten along any computation of $\cC$, so that it can be
represented by precisely one parameter. An occurrence of the binding
operator $\downarrow_r$ in $\varphi$ is represented in the product
automaton by an equality test $\mathrm{eq}(x_r)$. A positive
occurrence of a formula of the type $\uparrow_r$ is likewise
represented by an equality test $\mathrm{eq}(x_r)$, while a negative
occurrence of such a subformula is represented by a disequality test
$\tau(v_r)=\{x_r\}$.

Rather than recapitulating the constructions and reasoning
underlying~\cite[Theorem 15]{DemriS10}, we give below an extended
example that demonstrates the main ideas behind that result and helps
motivate the subsequent development in this paper.

\begin{exa}
  \label{example:fltl_reduction}
  Consider the flat $\FLTL$ formula
  \[\varphi \equiv true \oU (\downarrow_r \oX (\uparrow_r \wedge \oX \uparrow_r))\]
  and the four counter automata represented in
  Figure~\ref{fig:example_fltl_four_automata}. We will assume that for
  every 1-CA in this example the initial counter value is $c$.  We
  describe in detail a 1-CA with parameterised tests, denoted
  $\cC_1^{(\varphi)}$, that arises as the product of $\cC_1$ and
  $\varphi$.  We moreover explain how this definition changes if we
  replace $\cC_1$ with one of the remaining three 1-CA.

\textcolor{black}{
  To construct $\cC_1^{(\varphi)}$, the first step is to introduce a
  concrete representation of the syntax tree of $\varphi$, which we
  denote by $T_{\varphi}$ (see Figure~\ref{fig:example_fltl_tree}).
  This representation is convenient for distinguishing different
  occurrences of the same subformula within $\varphi$.  Each node of
  $T_{\varphi}$ is labelled with its \emph{address} (a word from
  $\{0,1\}^*$) together with an operator or atomic formula,
  corresponding to an occurrence of a subformula in $\varphi$. The
  address of the root node is $\epsilon$.  If a node has address $w$
  then its leftmost child (if it has a child) has address $w0$ and its second child (if it
  has two children) has address $w1$.  The operator label of every node is
  assigned in the obvious way, taking the outermost connective of the
  corresponding subformula.}

  \begin{figure}
  \centering
  \begin{minipage}{.5\textwidth}
    \centering
    \begin{tikzpicture}[every node/.style={scale=0.8}, every loop/.style={min distance=10mm,in=65,out=115,looseness=3}]
      \node[state] (v1) {$v_1$};
      \node[state] (v2) [right=1.2cm of v1] {$v_2$};
      \node[state] (v3) [right=1.2cm of v2] {$v_3$};
      \path[->]
      (v1) edge node [above] {$+0$} (v2)
      (v2) edge node [above] {$+0$} (v3)
      (v3) edge [loop above] node {$+1$} ();
    \end{tikzpicture}
    \caption*{$\cC_1$}
  \end{minipage}%
  \begin{minipage}{.5\textwidth}
    \centering
    \begin{tikzpicture}[every node/.style={scale=0.8}, every loop/.style={min distance=10mm,in=65,out=115,looseness=3}]
      \node[state] (v1) {$v_1$};
      \node[state] (v2) [right=1.2cm of v1] {$v_2$};
      \node[state] (v3) [right=1.2cm of v2] {$v_3$};
      \path[->]
      (v1) edge node [above] {$+1$} (v2)
      (v2) edge node [above] {$+0$} (v3)
      (v3) edge [loop above] node {$+0$} ();
    \end{tikzpicture}
    \caption*{$\cC_2$}
  \end{minipage}
  \begin{minipage}{.5\textwidth}
    \centering
    \begin{tikzpicture}[every node/.style={scale=0.8}, every loop/.style={min distance=10mm,in=65,out=115,looseness=3}]
      \node[state] (v1) {$v_1$};
      \node[state] (v2) [right=1.2cm of v1] {$v_2$};
      \node[state] (v3) [right=1.2cm of v2] {$v_3$};
      \path[->]
      (v1) edge node [above] {$+1$} (v2)
      (v2) edge node [above] {$+1$} (v3)
      (v3) edge [loop above] node {$+0$} ();
    \end{tikzpicture}
    \caption*{$\cC_3$}
  \end{minipage}%
  \begin{minipage}{.5\textwidth}
    \centering
    \begin{tikzpicture}[every node/.style={scale=0.8}, every loop/.style={min distance=10mm,in=65,out=115,looseness=3}]
      \node[state] (v1) {$v_1$};
      \node[state] (v2) [right=1.2cm of v1] {$v_2$};
      \node[state] (v3) [right=1.2cm of v2] {$v_3$};
      \path[->]
      (v1) edge node [above] {$+1$} (v2)
      (v2) edge node [above] {$+1$} (v3)
      (v3) edge [loop above] node {$+1$} ();
    \end{tikzpicture}
    \caption*{$\cC_4$}
  \end{minipage}
    \caption{The automata considered in Example~\ref{example:fltl_reduction}}
  \label{fig:example_fltl_four_automata}
\end{figure}

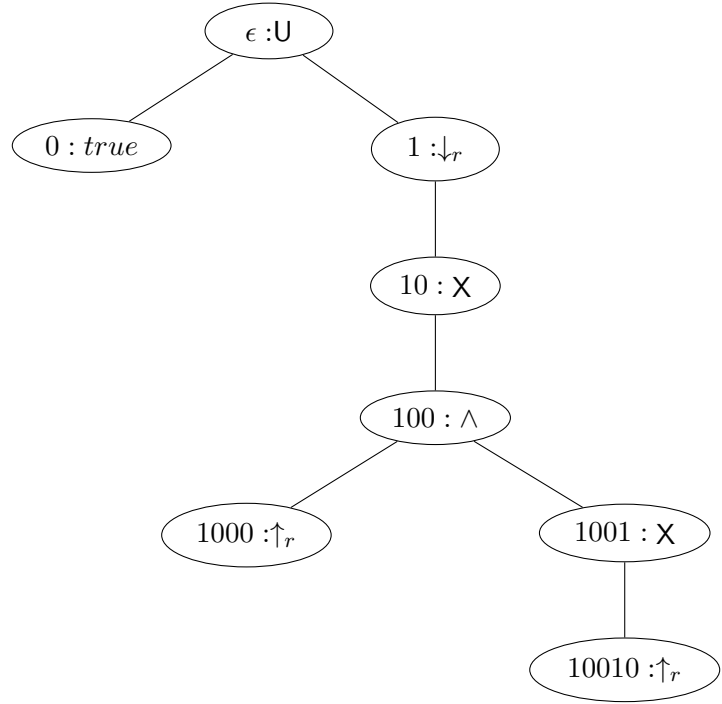
\begin{figure}
  \centering
  \begin{tikzpicture}
    \tikzstyle{occ} = [draw, ellipse, minimum width=1.7cm];
    \node[occ] (e) {$\epsilon : \oU$};
    \node[occ] (0) [below left=of e] {$0 : true$};
    \node[occ] (1) [below right=of e] {$1 : \downarrow_r$};
    \node[occ] (10) [below=of 1] {$10 : \oX$};
    \node[occ] (100) [below=of 10] {$100 : \wedge$};
    \node[occ] (1000) [below left=of 100] {$1000 : \uparrow_r$};
    \node[occ] (1001) [below right=of 100] {$1001 : \oX$};
    \node[occ] (10010) [below=of 1001] {$10010 : \uparrow_r$};
    \path[-]
    (e) edge node {} (0)
        edge node {} (1)
    (1) edge node {} (10)
    (10) edge node {} (100)
    (100) edge node {} (1000)
          edge node {} (1001)
    (1001) edge node {} (10010);
  \end{tikzpicture}
  \caption{The formula tree $T_\varphi$}
  \label{fig:example_fltl_tree}
\end{figure}

\textcolor{black}{
Next we introduce the notion of \emph{atoms} of $\varphi$ which are certain
sets of (occurrences of) subformulas of $\varphi$.  
Formally an atom $A$
is any subset of $\{0,1\}^*$
that satisfies the following conditions:\footnote{
In general there is also a condition for negation, but this is not relevant 
for the current simple example.}
  \begin{enumerate}
    \item If $w \in A$ and the node in $T_{\varphi}$ with address $w$ has 
label $\wedge$, then $w0,w1 \in A$.
    \item If $w \in A$ and the node in $T_{\varphi}$ with address $w$ has 
label $\mathsf{U}$, then $w0\in A$ or $w1\in A$.
    \item If $w \in A$ and the node in $T_{\varphi}$ with address $w$ has 
label $\downarrow_r$, then $w0 \in A$.
  \end{enumerate}
  The above conditions correspond to the intuition that an atom
  represents a set of subformulas of $\varphi$ that hold at a certain
  position in a data word.  We moreover define a transition relation
  between atoms by specifying that for atoms $A,A'$ we have
  $A \longrightarrow A'$ if the following conditions hold:
\begin{enumerate}
\item If $w \in A$ and the node in $T_{\varphi}$ with address $w$
  has label $\mathsf{X}$ then $w0 \in A'$.
\item If $w \in A$ and the node in $T_{\varphi}$ with address $w$
  has label $\mathsf{U}$ then either $w1 \in A$ or $w0 \in A$
  and $w\in A'$.
\item No atom $A''$ strictly included in $A'$ satisfies the preceding two
  conditions (with $A''$ in place of $A'$).
\end{enumerate}}

For the formula $\varphi$ at hand, $\{ \epsilon,0\}$ is an atom and the 
set of atoms reachable from this one under the transition relation
 is $\{ \{\epsilon,0\}, \{\epsilon,1,10\}, \{100,1000,1001\}, \{10010\}, \emptyset \}$, with the transition relation being given by:
  \begin{align*}
    \{\epsilon,0\} & \longrightarrow \{\epsilon,0\}\\
    \{\epsilon,0\} & \longrightarrow \{\epsilon,1,10\}\\
    \{\epsilon,1,10\} & \longrightarrow \{100,1000,1001\}\\
    \{100,1000,1001\} & \longrightarrow \{10010\}\\
    \{10010\} & \longrightarrow \emptyset
  \end{align*}

  Roughly speaking, the automaton $\cC_1^{(\varphi)}$ arises as the
  product of $\cC_1$ and the transition relation on atoms of
  $\varphi$.  This automaton is shown in
  Figure~\ref{fig:example_fltl_product}. It has an initial state $v_0$
  and auxiliary states $v_1^{aux},\ldots,v_6^{aux}$. All other states
  are of the form $\langle v,A \rangle$ where $v$ is a state of
  $\cC_1$ and $A$ is an atom of $\varphi$. From the initial state
  $v_0$ there is a nondeterministic choice between the two atoms
  having $\epsilon$ as an element. The choice of $\{\epsilon,1,10\}$
  means that $\downarrow_r \oX (\uparrow_r \wedge \oX \uparrow_r)$
  (the right side of $\varphi$) holds in the initial state $v_1$ of
  $\cC_1$, and the choice of $\{\epsilon,0\}$ means that this
  subformula will hold at some point in the future, i.e., in $v_2$ or
  $v_3$.  \textcolor{black}{ For any two consecutive edges
\[ \langle v_i,A \rangle \longrightarrow v_k^{aux} \longrightarrow
  \langle v_j,A' \rangle \] in the product automaton
$\cC_1^{(\varphi)}$, the label on the second edge is equal to the
label on the edge $(v_i,v_j)$ in automaton $\cC_1$, and the label on
the first edge is a test on the counter value.} There is one set of
final states, $F = \{\langle v_3,\emptyset \rangle\}$.\footnote{In
  general, there is a set of final states for each $\mathsf{U}$
  operator in the formula.}

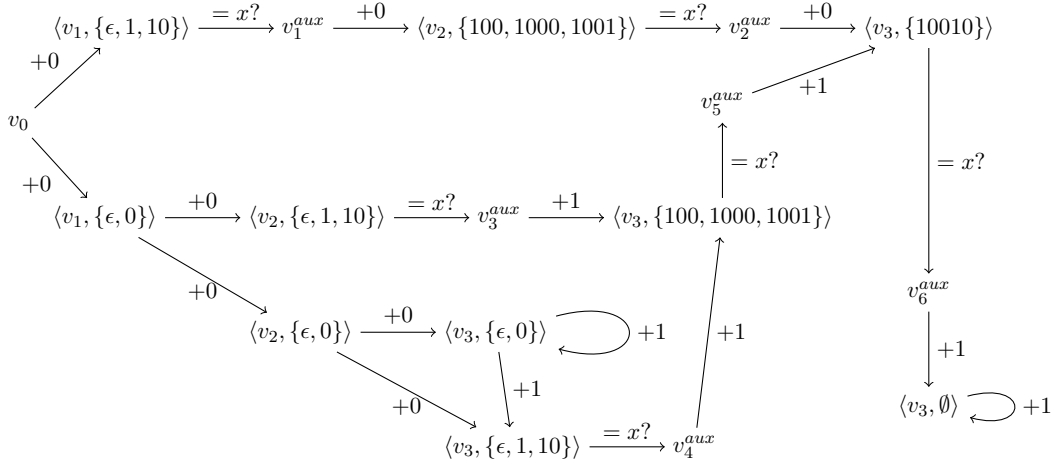
\begin{figure}
  \centering
  \begin{tikzpicture}[every node/.style={scale=0.8}]
    \tikzstyle{ps} = [draw, ellipse, minimum width=0.1cm];
    \node (v0) {$v_0$};
    \node (v1XeX1X10) [above right=0.8cm and 0.1cm of v0] {$\langle v_1, \{ \epsilon, 1, 10\} \rangle$};
    \node (v1aux) [right=of v1XeX1X10] {$v_1^{aux}$};
    \node (v2X100X1000X1001) [right=of v1aux] {$\langle v_2, \{ 100,1000,1001\} \rangle$};
    \node (v2aux) [right=of v2X100X1000X1001] {$v_2^{aux}$};
    \node (v3X10010) [right=of v2aux] {$\langle v_3, \{ 10010 \} \rangle$};
    \node (v1XeX0) [below right=0.8cm and 0.1cm of v0] {$\langle v_1, \{ \epsilon,0\} \rangle$};
    \node (v2XeX1X10) [right=of v1XeX0] {$\langle v_2, \{ \epsilon, 1,10 \} \rangle$};
    \node (v3aux) [right=of v2XeX1X10] {$v_3^{aux}$};
    \node (v3X100X1000X1001) [right=of v3aux] {$\langle v_3, \{100,1000,1001\} \rangle$};
    \node (v2XeX0) [below right=of v1XeX0] {$\langle v_2, \{ \epsilon,0\} \rangle$};
    \node (v3XeX0) [right=of v2XeX0] {$\langle v_3, \{ \epsilon,0\} \rangle$};
    \node (v3XeX1X10) [below right=of v2XeX0] {$\langle v_3, \{\epsilon,1,10\} \rangle$};
    \node (v4aux) [right=of v3XeX1X10] {$v_4^{aux}$};
    \node (v5aux) [above=of v3X100X1000X1001] {$v_5^{aux}$};
    \node (v6aux) [below=3cm of v3X10010] {$v_6^{aux}$};
    \node (v3Xo) [below=of v6aux] {$\langle v_3, \emptyset \rangle$};
    \path[->]
    (v0) edge node [above left] {$+0$} (v1XeX1X10)
         edge node [below left] {$+0$} (v1XeX0)
    (v1XeX1X10) edge node [above] {$=x?$} (v1aux)
    (v1aux) edge node [above] {$+0$} (v2X100X1000X1001)
    (v2X100X1000X1001) edge node [above] {$=x?$} (v2aux)
    (v2aux) edge node [above] {$+0$} (v3X10010)
    (v1XeX0) edge node [above] {$+0$} (v2XeX1X10)
             edge node [below] {$+0$} (v2XeX0)
    (v2XeX1X10) edge node [above] {$=x?$} (v3aux)
    (v3aux) edge node [above] {$+1$} (v3X100X1000X1001)
    (v2XeX0) edge node [above] {$+0$} (v3XeX0)
             edge node [below] {$+0$} (v3XeX1X10)
    (v3XeX0) edge [loop right] node {$+1$} ()
             edge node [right] {$+1$} (v3XeX1X10)
    (v3XeX1X10) edge node [above] {$=x?$} (v4aux)
    (v4aux) edge node [right] {$+1$} (v3X100X1000X1001)
    (v3X100X1000X1001) edge node [right] {$=x?$} (v5aux)
    (v5aux) edge node [below] {$+1$} (v3X10010)
    (v3X10010) edge node [right] {$=x?$} (v6aux)
    (v6aux) edge node [right] {$+1$} (v3Xo)
    (v3Xo) edge [loop right] node {$+1$} ();
  \end{tikzpicture}
  \caption{The product automaton $\cC_1^{(\varphi)}$ of $\cC_1$ and $\varphi$.}
  \label{fig:example_fltl_product}
\end{figure}

Let us first follow the path from
$\langle v_1, \{ \epsilon,1,10\} \rangle$. The transition to the
auxiliary state $v_1^{aux}$ tests whether the current counter value,
that is $c$, is equal to the parameter $x$. This equality test
corresponds to node $1$ in $T_\varphi$. Next, $\cC_1^{(\varphi)}$
simulates a transition from $v_1$ to $v_2$ in $\cC_1$. The transitions
to auxiliary states $v_2^{aux}$ and $v_6^{aux}$ include further tests
for equality with $x$, which correspond to nodes $1000$ and $10010$ in
$T_\varphi$, respectively. Clearly if we set $x=c$, there is a valid
computation starting from $(v_0,c)$ over this path which ends up
visiting the final state $\langle v_3,\emptyset \rangle$ infinitely
many times.

The paths starting at $\langle v_1,\{\epsilon,0\}\rangle$ all
correspond to cases where
$\downarrow_r \oX (\uparrow_r \wedge \oX \uparrow_r)$ does not hold in
$v_1$, but only at a later time in $v_2$ or $v_3$. It is easy to see
that there is no value for $x$ that allows reachability of the final
state along any of these paths.

Recall the 1-CA $\cC_2$, $\cC_3$ and $\cC_4$ from
Figure~\ref{fig:example_fltl_four_automata}. In the 1-CA
$\cC_2^{(\varphi)}$, which is constructed in the same way as
$\cC_1^{(\varphi)}$ but with transition labels from $\cC_2$ rather
than $\cC_1$, there is a valid computation starting in $(v_0,c)$ over
the path that goes through $v_3^{aux}$ and $v_5^{aux}$ and ends up
visiting $\langle v_3,\emptyset\rangle$ infinitely many times. This
means that the right side of $\varphi$ becomes true in $\cC_2$ after
one step. Similarly, in $\cC_3^{(\varphi)}$, there is a valid
computation over any path that goes through $v_4^{aux}$ and
$v_5^{aux}$ to $\langle v_3,\emptyset\rangle$, which means that the
right side of $\varphi$ becomes true in $\cC_3$ after two or more
steps. Finally, in $\cC_4^{(\varphi)}$, the state
$\langle v_3,\emptyset\rangle$ can never be reached, since there is no
computation from $(v_1,c)$ in $\cC_4$ that satisfies $\varphi$.
  
If $\varphi$ featured any negated atomic formulas of the form
$\downarrow_{r'} \varphi'$, auxiliary states with disequality tests
would be needed in the product automaton. For a general description of
how to construct the product automaton and a proof, see
\cite{DemriS10}.
\end{exa}

Note that the definition of 1-CA with parameterised tests
in~\cite{DemriS10} includes parameterised equality and disequality tests
(as in the present paper) together with parameterised inequality tests,
i.e., testing whether the counter value is less than or greater than the
value of a parameter.  However, it is clear from the details of the
reduction that only equality and disequality tests are needed, and
thus we do not consider inequality tests in this paper.
%%%not needed any more, explained in detail in previous section
%Note also
%that in the previous section we defined 1-CA to have
%equality tests on edges and disequality tests on states.  On the other
%hand, the 1-CA considered in \cite{DemriS10} have both
%kinds of tests on edges and allow multiple edges between the same pair
%of states. It is easy to see that both models are equivalent with
%respect to reachability, i.e., there are reductions in both directions
%between reachability problems in the two models.

%%% Local Variables:
%%% mode: latex
%%% TeX-master: "Main-Revised.tex"
%%% End:

\subsection{Presburger Arithmetic}
\emph{Presburger arithmetic} is the first-order logic over the
structure $\langle \mathbb{Z}, +, <, 0, 1 \rangle$, where $+$ and $<$
are the standard addition and ordering on integers. Presburger
arithmetic is known to be decidable~\cite{Pres29}.  Using shorthand
notation, we can assume that the atomic formulas of Presburger
arithmetic are equalities or inequalities between linear polynomials
with integer coefficients.

%%% Local Variables:
%%% mode: latex
%%% TeX-master: "Main-Revised.tex"
%%% End:

\section{Normal Form for Paths}
\label{sec:normal}
In this section, we show that any valid finite computation of a 1-CA
$\cC=(V,E,\lambda,\tau)$ can be rewritten to a normal form whose shape
only depends on the automaton and such that the initial and final
configuration of the computation are preserved. Informally, any such
computation can be described as a sequence of ``take this transition''
and ``take this cycle $k$ times''. We show that the maximum length of
a description of this kind is independent of the
original computation.  \textcolor{black}{Such a description is similar
  in spirit to the semilinear path schemes described
  in~\cite{LerouxS05}.}

\textcolor{black}{We give a brief overview of the technical development
  below.  The first step (Lemma~\ref{lem:eq_test_isolate}) is to bound
  the number of equality tests along a minimum-length computation
  between two configurations of a 1-CA.  Thereafter we focus on
  computations that are free of equality tests.  To obtain a succinct
  representation of such computations we define a rewriting system
  that reorders computations by gathering together in the same place
  executions of the same simple cycle; we moreover introduce a
  compressed representation of iterated simple cycles, leading to the
  notion of \emph{folded paths}.  Lemmas~\ref{lem:rewrite_sound}
  and~\ref{lem:rewrite_terminates} show that the rewriting rules are
  sound (i.e., preserve validity of computations) and terminating.  We
  then concentrate on bounding the length of folded paths which cannot
  be further rewritten.  To this end we identify a set of ``critical''
  configurations that block application of the rewriting rules, and we
  bound the number of such configurations
  (Lemma~\ref{lem:no_gather_obstruction}).  This leads to an upper
  bound on the length of a folded path that cannot be rewritten
  (Lemmas~\ref{lem:no_rule_count_cycles}
  and~\ref{lem:nf_eq_free_comp}). Finally the whole analysis,
  including equality tests, is summarised in Theorem~\ref{th:nf_path}
  which gives the required upper bound on the length of folded paths.}

In the rest of this section we consider a fixed 1-CA
$\cC=(V,E,\lambda,\tau)$.  First we show that, without loss of
generality, any computation in $\cC$ can be broken down into a small
number of segments that do not contain any transitions with equality
tests. The idea is that any segment between two identical equality
tests can be omitted.

\begin{lem}[Equality-test isolation]\label{lem:eq_test_isolate}
Let $\pi$ be a valid finite computation from $(v,c)$ to $(v',c')$. Then there
exists a path $\gamma$ such that $\gamma(c)$ is a valid computation from $(v,c)$ to $(v',c')$
and $\gamma$ is of the form $\gamma=\gamma_0e_1\gamma_1e_2\cdots e_n\gamma_n$,
where $e_i$ is an edge with an equality test, $\gamma_i$ is a path without equality tests
and $n\leqslant|E|$.
\end{lem}

\begin{proof}
Let $\pi'$ be the shortest % (in length)
valid computation from $(v,c)$ to $(v',c')$. %We know it exists
%because $\pi$ is such a  computation.
We can decompose it as
\[
\pi'=(v,c)\xrightarrow{\gamma_0}(v_1,c_1)\xrightarrow{e_1}(v_1',c_1)\xrightarrow{\gamma_1}(v_2,c_2)
\cdots (v_n',c_n)\xrightarrow{\gamma_n}(v',c')
\]
where for every $i$, $\gamma_i$ is a path without any equality tests and $e_i=(v_i,\mathrm{eq}(c_i),v_i')\in E$
is an equality test. Then clearly $\pi'=\gamma(c)$ where
\[\gamma=\gamma_0e_1\gamma_1e_2\cdots e_n\gamma_n.\]
Assume for a contradiction that $n>|E|$. Then by the pigeonhole principle, there exists $i<j$ such that $e_i=e_j$.
  But since $\pi'$ is a valid computation, the two transitions $(v_i,c_i)\xrightarrow{\mathrm{eq}(c_i)}(v_i',c_i)$
and  $(v_j,c_j)\xrightarrow{\mathrm{eq}(c_i)}(v_j',c_j)$ are the same and $(v_i,c_i)=(v_j,c_j)$.
Thus we can delete part of the computation and define
\[\gamma'=\gamma_0e_1\gamma_1\cdots e_i\gamma_j e_{j+1}\cdots e_n\gamma_n.\]
Then $\gamma'(c)$ is a valid computation from $(v,c)$ to $(v',c')$ and is shorter than $\pi'$, which
is a contradiction.
\end{proof}

We need to introduce some terminology to formalise our notion of
normal form.  
%Given a state $v$, $\simpcycle{v}$ denotes the set of
%\emph{equality-free simple cycles} starting at $v$.  We moreover write
%$\simpcyclepos{v}$ for the set of equality-free \emph{positive} simple
%cycles starting at $v$ and likewise $\simpcycleneg{v}$ for the
%negative such cycles.  
Write $\simplecycleall$ for the set of all equality-free simple
cycles in $\cC$.  We moreover denote by
$\simpcycleposall \subseteq \simplecycleall$ the set of equality-free
simple cycles that have positive weight and likewise
by $\simpcyclenegall \subseteq \simplecycleall$ the set of cycles with
negative weight.

The \emph{cycle alphabet}, denoted $C$, consists of symbols of the
form $\cyclemark{\omega^k}$ where $\omega\in\simplecycleall$ and
$k\in\N$. Note that this alphabet is infinite. Also note that
$\cyclemark{\omega^k}$ is a single symbol, underlined to 
distinguish it from the cycle $\omega^k$, which consists of
$\lvert \omega \rvert k$ symbols from $E$.  For convenience,
we use $\cyclemark{\omega}$ as shorthand for $\cyclemark{\omega^1}$. We
naturally define the start and end of symbol $\cyclemark{\omega^k}$ by
the start of $\omega$:
$\pathstart{\cyclemark{\omega^k}}=\pathend{\cyclemark{\omega^k}}=\pathstart{\omega}$.

A \emph{folded path} $\chi$ is a word over the alphabet $E\cup C$:
$\chi=s_1\cdots s_n$ such that $\pathend{s_i}=\pathstart{s_{i+1}}$ for
every $i<n$. We also define the natural unfolding folded paths via
a monoid homomorphism
$\operatorname{unfold}: (E\cup C)^*\rightarrow E^*$ such that
$\pathunfold{e}=e$ for $e\in E$ and
$\pathunfold{\cyclemark{\omega^k}}=\omega^k$ for
$\cyclemark{\omega^k} \in C$. The weight of a folded path is the
weight of its unfolding.

For the rest of this section we fix an initial counter value
$c \in \N$ and we only consider computations starting at $c$ that do
not feature equality tests.  We refer to a folded path $\chi$ as being
\emph{valid} if $\pathunfold{\chi}(c)$ is a valid computation.

Define the following nondeterministic rewriting system on folded
paths.  Each rule of the system has a name, a pattern to match
against, a condition that must be satisfied for the rule to apply, and
the result of the rule. We denote by $\chi\leadsto\chi'$ the fact that
$\chi$ rewrites to $\chi'$.

\begin{center}
\begin{tabular}{@{}cccp{8.2cm}@{}}
\textbf{Rule}&\textbf{Pattern}&\textbf{Result}&\textbf{Condition}\\
\hline
\texttt{fold} & $\psi\omega\phi$ & $\psi\cyclemark{\omega}\phi$ & $\omega$ is a simple cycle of nonzero weight.\\
\texttt{simplify} & $\psi\rho\phi$ & $\psi\phi$ & Nonempty $\rho$, $\pathweight{\pathunfold{\rho}}=0$ and $\pathend{\psi}=\pathstart{\phi}$.\\
\texttt{gather\textsuperscript{+}} & $\psi\cyclemark{\omega^k}\rho\cyclemark{\omega^\ell}\phi$
    & $\psi\cyclemark{\omega^{k+1}}\rho\cyclemark{\omega^{\ell-1}}\phi$
    & Result is valid, $\omega$ is a positive simple cycle and $\ell>0$.\\
\texttt{gather\textsuperscript{-}} & $\psi\cyclemark{\omega^k}\rho\cyclemark{\omega^\ell}\phi$
    & $\psi\cyclemark{\omega^{k-1}}\rho\cyclemark{\omega^{\ell+1}}\phi$
    & Result is valid, $\omega$ is a negative simple cycle and $k>0$.\\
\end{tabular}
\end{center}

\begin{lem}[Soundness]\label{lem:rewrite_sound}
  If $\chi$ is a valid folded path that rewrites to $\chi'$, then
  $\chi'$ is also valid.  Furthermore, $\chi$ and $\chi'$ start and
  end at the same state and
  $\pathweight{\pathunfold{\chi}} = \pathweight{\pathunfold{\chi'}}$.
\end{lem}

\begin{proof}
This is easily checked for each rule:
\begin{itemize}
  \item\texttt{fold}: Clearly $\pathunfold{\chi} = \pathunfold{\chi'}$.
\item\texttt{simplify}: First note that the result is well-formed because of the condition
  on start and end. The unfolding of the first part ($\psi$) of the path is unchanged, so it remains valid and with the same starting state. Since the second part of the path ($\rho$) has weight $0$, the counter value is the same at the beginning and end of
$\rho$, so the unfolding of the third part ($\phi$) stays the same, and thus valid with the same end state. The weight of the unfolded path remains unchanged as the removed part $\rho$ has weight $0$.
\item\texttt{gather\textsuperscript{$\pm$}}: The condition ensures the result is valid. The start and end state clearly do not change, and neither does the weight, since $\pathunfold{\chi'}$ contains the same edges as $\pathunfold{\chi}$, only in a different order. \qedhere
%\item\texttt{gather\textsuperscript{-}}: The condition ensures the result is valid.
\end{itemize}
%The fact that the start, end and weight are the same before and after is trivial to check for each rule.
\end{proof}

\begin{lem}[Termination]\label{lem:rewrite_terminates}
There are no infinite chains of rewriting.
\end{lem}

%A proof of Lemma~\ref{lem:rewrite_terminates} can be found in Appendix~\ref{sec:appendix}. 
%Appendix~\ref{sec:appendix}.
%Here we give an informal explanation.
%The first thing to notice is that the length of a folded path (over
%alphabet $E \cup C$) never increases after a rewriting operation. The second thing is that the
%length of a folded path over $E$ (i.e., ignoring symbols from $C$) never
%increases either.  Since rule \texttt{simplify} decreases the length,
%it can only be applied finitely many times.  Similarly, rule
%\texttt{fold} decreases the length over $E$ because it replaces a
%symbol from $E$ by one from $C$. Rules
%\texttt{gather\textsuperscript{$\pm$}} are more difficult to analyse
%because they only reorder the path by replacing symbols from $C$. But
%as it can be seen, a symbol $\cyclemark{\omega}$, where $\omega$ is a
%positive cycle, can only move left, and similarly a negative cycle can
%only move right. Intuitively, this process must be finite because once
%a positive (negative) cycle reaches the leftmost (rightmost) position,
%it cannot move anymore.

\begin{proof}
First we give an informal explanation.
The first thing to notice is that the length of a folded path (over
alphabet $E \cup C$) never increases after a rewriting operation. The second thing is that the
length of a folded path over $E$ (i.e., ignoring symbols from $C$) never
increases either.  Since rule \texttt{simplify} strictly decreases the length,
it can only be applied finitely many times.  Similarly, rule
\texttt{fold} strictly decreases the length over $E$ because it replaces a
symbol from $E$ by one from $C$. Rules
\texttt{gather\textsuperscript{$\pm$}} are more difficult to analyse
because they only reorder the path by replacing symbols from $C$. But
notice that a symbol $\cyclemark{\omega}$, where $\omega$ is a
positive cycle, can only move left, and similarly a negative cycle can
only move right. Intuitively, this process must be finite because once
a positive (negative) cycle reaches the leftmost (rightmost) position,
it cannot move anymore.

Formally, we will define a valuation over folded paths and show that it
  decreases after each application of a rule. First, for any folded
  path $\chi$ and any given simple cycle $\omega$, define the
  \emph{$\omega$-projection} $p_\omega(\chi)$ of $\chi$ to be the subword
  consisting only of symbols of the form $\cyclemark{\omega^k}$:
\[p_\omega(e\chi)=p_\omega(\chi)\text{ if }e\in E
\qquad p_\omega(\cyclemark{\omega^k}\chi)=\cyclemark{\omega^k}p_\omega(\chi)
\qquad p_\omega(\cyclemark{\theta^k}\chi)=p_\omega(\chi)\text{ if }\theta\neq\omega.\]
For any folded path $\chi$, define:
\[\llparenthesis\chi\rrparenthesis=(|\chi|,|\chi|_E,\sigma(\chi)),\quad\text{ where }\quad
\sigma(\chi)=\sum_{\omega\in\simplecycleall}\sigma_\omega(p_\omega(\chi)) \, , \]
$|\chi|$ is the word length of $\chi$ (over alphabet $E\cup C$),
$|\chi|_E$ is the word length of $\chi$ only counting symbols in $E$, and $\sigma_\omega(p_\omega(\chi))$
is defined as follows:
\[\sigma_\omega\left(\cyclemark{\omega^{k_1}}\cyclemark{\omega^{k_2}}\cdots\cyclemark{\omega^{k_n}}\right)
=\begin{cases}
\sum_{i=1}^nik_i&\text{if }\pathweight{\omega}>0\\
0&\text{if }\pathweight{\omega}=0\\
\sum_{i=1}^n(n+1-i)k_i)&\text{if }\pathweight{\omega}<0.
\end{cases}\]

We will now show that $\llparenthesis\chi\rrparenthesis$ decreases in lexicographic
order each time a rule is applied.
In the case of rule \texttt{fold}, if $|\omega| \geq 2$ then clearly $|\chi|$ decreases because we replace
several symbols with just one. If $|\omega|=1$ then $|\chi|$ stays constant but
$|\chi|_E$ decreases by one because we replace one symbol from $E$ by one symbol from $C$.
Similarly, rule \texttt{simplify} decreases $|\chi|$ because we remove a nonzero-length subpath.
Since rules \texttt{gather\textsuperscript{+}} and \texttt{gather\textsuperscript{-}}
are symmetric, we only consider \texttt{gather\textsuperscript{+}}. Note
that the rule does not change $|\chi|$ or $|\chi|_E$ because it only 
%% (changed this part because technically, the path is -reordered-, but symbols are -replaced-) 
%reorders symbols, so
replaces symbols from $C$ with different symbols from $C$, so
we are only concerned with $\sigma(\chi)$.

Assume the rule rewrites $\psi\cyclemark{\omega^k}\rho\cyclemark{\omega^\ell}\phi$
into $\psi\cyclemark{\omega^{k+1}}\rho\cyclemark{\omega^{\ell-1}}\phi$. First note
that if $\theta\neq\omega$ is a simple cycle, then the $\theta$-projection is the same before
and after the rule because the rule does not %reorder
replace any symbols of the form $\cyclemark{\theta^k}$, so $\sigma_\theta$ does not change.
%and thus:
%\[\sigma_\theta\left(\psi\cyclemark{\omega^k}\rho\cyclemark{\omega^\ell}\phi\right)=
%\sigma_\theta\left(\psi\cyclemark{\omega^{k+1}}\rho\cyclemark{\omega^{\ell-1}}\phi\right).\]
The case of $\sigma_\omega$ is slightly more involved and we need to introduce some notations:
\[p_\omega(\psi)=\cyclemark{\omega^{u_1}}\cdots\cyclemark{\omega^{u_n}},
\qquad p_\omega(\rho)=\cyclemark{\omega^{u_{n+2}}}\cdots\cyclemark{\omega^{u_m}},
\qquad p_\omega(\phi)=\cyclemark{\omega^{u_{m+2}}}\cdots\cyclemark{\omega^{u_q}}\]
and
\[u_{n+1}=k,\quad u_{m+1}=\ell,\quad u'_{n+1}=k+1,\quad u'_{m+1}=\ell-1,\quad u'_i=u_i\text{ if }i\neq n+1,m+1.\]
Then we can observe that:
\begin{align}
\sigma_\omega\left(p_\omega\left(\psi\cyclemark{\omega^k}\rho\cyclemark{\omega^\ell}\phi\right)\right)
    &=\sigma_\omega\left(\cyclemark{\omega^{u_1}}\cdots\cyclemark{\omega^{u_q}}\right)
    =\sum_{i=1}^q iu_i,\label{eq_sigma_before}\\
\sigma_\omega\left(p_\omega\left(\psi\cyclemark{\omega^{k+1}}\rho\cyclemark{\omega^{\ell-1}}\phi\right)\right)
    &=\sigma_\omega\left(\cyclemark{\omega^{u'_1}}\cdots\cyclemark{\omega^{u'_q}}\right)
    =\sum_{i=1}^q iu'_i\label{eq_sigma_after}.
\end{align}
Thus:
\begin{align*}
\eqref{eq_sigma_before}-\eqref{eq_sigma_after}
&=\sum_{i=1}^q i(u_i-u'_i)\\
&=(n+1)(u_{n+1}-u'_{n+1})+(m+1)(u_{m+1}-u'_{m+1})\\
&=-(n+1)+(m+1)\\
&>0\text{ because }m>n.
\end{align*}
Thus $\sigma_\omega(\chi)$ decreases after the rule is applied and thus $\sigma(\chi)$
also decreases.
\end{proof}

\begin{lem}[Size of cycle-free subpaths]\label{lem:no_rule_small_E_words}
If $\psi\rho\phi$ is a folded path such that $\rho\in E^*$ and no rewriting rule applies, then $|\rho|<|V|$.
\end{lem}

\begin{proof}
Assume the contrary: if $\rho$ only consists of edges and has length $\geqslant|V|$,
then some state is repeated in the state sequence of $\rho$. Thus $\rho$ contains a cycle
and thus a simple cycle. So rule \texttt{fold} applies if the cycle has nonzero
weight, or rule \texttt{simplify} applies if it has weight zero.
\end{proof}

\textcolor{black}{The next lemma analyses situations in which the
  pattern of one of the rules \texttt{gather\textsuperscript{+}} and
  \texttt{gather\textsuperscript{-}} matches a factor of a folded word,
  but application of the rule leads to an invalid computation.  The
  idea is to identify a set of so-called critical configurations which
  can potentially prevent application of one of these two rules and then to
  bound the number of such critical configurations.  As we observe below, both
  rules are sound with respect to the requirement that
  counter values be nonnegative and can only cause a computation to
  become invalid through the presence of disequalilty tests.}

\textcolor{black}{ Given a state $v$ of $\cC$, we define a set
  $B^+(v)$ of \emph{critical values} for positive cycles and a set
  $B^{-}(v)$ of critical values for negative cycles.  These sets
  represent valid configurations $(v,c)$ from which some simple cycle
  cannot be executed due to a disequality test.  Formally, for
  $S \subseteq \Z$ and $x \in \Z$, write $S-x$ to denote
  $\{y-x \ |\ y \in S\}$; then we define $B^+(v)$ to be the union of
  the sets $\tau(\pathend{\gamma})-\pathweight{\gamma}$ for $\gamma$ a
  non-empty prefix of some positive cycle starting at $v$.  Likewise
  we define $B^{-}(v)$ to be the union of the sets
  $\tau(\pathend{\gamma})-\pathweight{\gamma}$ for $\gamma$ a
  non-empty prefix of some negative cycle starting at $v$.}

\begin{lem}[Obstructions in irreducible paths with cycles]\label{lem:no_gather_obstruction}
  Let $\omega$ be a positive (resp.\ negative) cycle and assume that
  rule \texttt{gather\textsuperscript{+}} (resp.\
  \texttt{gather\textsuperscript{-}}) does not apply to
  $\psi\cyclemark{\omega^k}\rho\cyclemark{\omega^\ell}\phi$ (which we
  assume is valid and $k,\ell>0$) for this particular pattern.  Then
  there exists a (potentially empty) prefix $\mu$ of $\rho$ such that
  $\pathunfold{\psi\cyclemark{\omega^k}\mu}(c)$ has the form
  $(v,c)\longrightarrow^*(v',c')$ where $c'$ is critical for $v'$ for
  positive (resp.\ negative) cycles, \emph{i.e.} $c'\in B^+(v')$
  (resp.\ $c'\in B^-(v')$). Furthermore $B^+(v')$ and $B^-(v')$ only
  depend on the automaton and
\[|B^+(v')|\leqslant|\simpcycleposall|\sum_{u\in V}|\tau(u)|\quad\text{ and }\quad
|B^-(v')|\leqslant|\simpcyclenegall|\sum_{u\in V}|\tau(u)|. \]
\end{lem}

\begin{proof}
We first show the result for positive cycles.
Let $\pi=\pathunfold{\psi\cyclemark{\omega^k}\rho\cyclemark{\omega^\ell}\phi}(c)$
and $\pi'=\pathunfold{\psi\cyclemark{\omega^{k+1}}\rho\cyclemark{\omega^{\ell-1}}\phi}(c)$.
To make things slightly easier to understand, note that:
\begin{align*}
  \pi&=[\pathunfold{\psi}\omega^k\pathunfold{\rho}\omega\omega^{\ell-1}\pathunfold{\phi}](c)\\
  \pi'&=[\pathunfold{\psi}\omega^k\omega\pathunfold{\rho}\omega^{\ell-1}\pathunfold{\phi}](c).
\end{align*}
Since $\pathunfold{\rho}\omega$ and $\omega\pathunfold{\rho}$ have the same weight,
it is clear that the first ($\pathunfold{\psi}\omega^k$) and last ($\omega^{\ell-1}\pathunfold{\phi}$)
parts of the computation are the same in $\pi$ and $\pi'$, i.e., they have the same counter values.
Consequently, if they are valid in $\pi$,
the same parts are also valid in $\pi'$. Since by the hypothesis \texttt{gather\textsuperscript{+}} does not apply,
$\pi'$ is invalid.
So there must be an obstruction $(u,d)$ in the middle part ($\omega\pathunfold{\rho}$)
of $\pi'$. There are two possibilities.

The first case is when the obstruction $(u,d)$ is in the $\pathunfold{\rho}$ part of $\pi'$.
Note that $d = c^* + \pathweight{\omega}$, where $(u,c^*)$ is the corresponding configuration in the
$\pathunfold{\rho}$ part of $\pi$. Since $\omega$ is a positive cycle,
$d > c^*$ cannot be negative (since $(u,c^*)$ occurs in $\pi$, which is valid).
Since we assumed that all computations are free of equality tests, the obstruction must be because of a disequality,
i.e., it must be that $d=c^*+\pathweight{\omega}\in\tau(u)$. Thus $c^*\in\tau(u)-\pathweight{\omega}$
and $c^*$ is critical for $u$.
Then there exists a prefix $\mu$ of $\rho$
such that $\pathunfold{\psi\cyclemark{\omega^k}\mu}(c)=(v,c)\longrightarrow^*(u,c^*)$
and this shows the result.

The second case is when $(u,d)$ is in the $\omega$ part of the middle part ($\omega\pathunfold{\rho}$)
of $\pi'$. Again, it is impossible that the counter value $d$ be negative. Indeed,
remember that $\omega$ is a positive cycle and $k>0$, thus
\begin{align*}
  \pi'&=[\pathunfold{\psi}\omega^{k+1}\pathunfold{\rho}\omega^{\ell-1}\pathunfold{\phi}](c)\\
       &=[\pathunfold{\psi}\omega^{k-1}\omega\omega\pathunfold{\rho}\omega^{\ell-1}\pathunfold{\phi}](c)\\
  &=(v,c)\xtransrightarrow{\pathunfold{\psi}\omega^{k-1}}(v_1,c_1)
    \xtransrightarrow{\omega}(v_1,c_2)\xtransrightarrow{\omega}(v_1,c_3)
    \xtransrightarrow{\pathunfold{\rho}\omega^{\ell-1}\pathunfold{\phi}}(v'',c'').
\end{align*}
We already argued that 
$(v,c)\longrightarrow^*(v_1,c_2)$ is valid,
so in particular $(v_1,c_1)\xtransrightarrow{\text{ }\omega\text{ }}(v_1,c_2)$ is valid. Note that
the obstruction is in the second iteration of $\omega$: $(v_1,c_2)\xtransrightarrow{\text{ }\omega\text{ }}(v_1,c_3)$.
Since $\omega$
is a positive cycle, $c_2>c_1$. Note that initially the cycle $\omega$ was feasible (with the counter not going negative)
starting with a lower counter value ($c_1$) so the counter cannot possibly become negative on the
second iteration starting with a higher counter value ($c_2$). Thus, again, the obstruction
happens because of a disequality. That is, we can write
$\omega=\gamma\gamma'$ such that:
\begin{align*}
\pi'&=(v,c)\xtransrightarrow{\pathunfold{\psi}\omega^k}(v_1,c_2)
  \xtransrightarrow{\text{ }\gamma\text{ }}(u,d)\xtransrightarrow{\text{ }\gamma'\text{ }}(v_1,c_3)
    \xtransrightarrow{\pathunfold{\rho}\omega^{\ell-1}\pathunfold{\phi}}(v'',c'')\\
\end{align*}
and the obstruction happens because $d\in\tau(u)$. Note however that $d=c_2+\pathweight{\gamma}$
and thus $c_2\in\tau(u)-\pathweight{\gamma}$. In this case, $c_2$ is critical for $v_1$.
Choose $\mu$ to be the empty word,
so that $\pathunfold{\psi\cyclemark{\omega^k}\mu}(c)=(v,c)\longrightarrow^*(v_1,c_2)$
to show the result.

Observe that the definition of critical values only depends on the
automaton itself. Furthermore, the size of $B^+(v)$
can easily be bounded. Indeed, there are $|\simpcycleposall|$ positive
simple cycles, for each such cycle its non-empty prefixes all end in
different states, and a prefix ending in a state $u$ contributes
$|\tau(u)|$ elements to $B^+(v)$.  It follows that
$|B^+(v)| \leq |\simpcycleposall|\sum_{u\in V}|\tau(u)|$.

The proof is exactly the same in the negative case except for one detail.
This time we move negative cycles to the right so that the middle part of $\pi'$
  ($\pathunfold{\rho}\omega$) can only get higher counter values than the middle part of $\pi$
  ($\omega \pathunfold{\rho}$), as in the positive case.
\end{proof}

\textcolor{black}{The following lemma is a step towards bounding the
  length of a folded path to which no rewriting rule applies.  We use this lemma
to obtain such a bound in Lemma~\ref{lem:nf_eq_free_comp}.}

\begin{lem}[Length of irreducible paths]\label{lem:no_rule_count_cycles}
Let $\chi$ be a folded path such that no rewriting rule applies on $\chi$.
Let $Y=\simpcycleposall$ or $Y=\simpcyclenegall$.
Then for every $\omega\in Y$, the number of symbols in $\chi$ of
the form $\cyclemark{\omega^\cdot}$ (the exponent does not matter) is bounded by
\[|V||Y|\Big(1+\sum_{v\in V}|\tau(v)|\Big).\]
\end{lem}

\begin{proof}
Without loss of generality, we show the result for $X=\simpcycleposall$.
First note that if $\cyclemark{\omega^k}$ appears in $\chi$ and no rule applies,
then $k>0$, otherwise we could apply \texttt{simplify} to remove $\cyclemark{\omega^0}$.
We can thus decompose the path as:
\[\chi=\phi_0\cyclemark{\omega^{k_1}}\phi_1\cyclemark{\omega^{k_2}}\phi_2\cdots\phi_{n-1}\cyclemark{\omega^{k_n}}\phi_n\]
where $k_i>0$ and $\phi_i$ does not contain any $\cyclemark{\omega^\cdot}$ symbol.
Since no rule applies, by Lemma~\ref{lem:no_gather_obstruction}, there exist prefixes
$\mu_1,\mu_2,\ldots,\mu_{n-1}$ of $\phi_1,\phi_2,\ldots,\phi_{n-1}$ respectively, such that for each $i$:
\[(v,c)\xtransrightarrow{\phi_0\cyclemark{\omega^{k_1}}\phi_1\cdots\phi_{i-1}\cyclemark{\omega^{k_i}}\mu_i}(v_i,c_i)
\quad\text{where}\quad c_i\in B^+(v_i).\]
Assume for a contradiction that there is a repeated configuration among the $(v_i,c_i)$.
Then there exists $i<j$ such that $v_i=v_j$ and $c_i=c_j$.
Let $\phi_i=\mu_i\rho$ and $\phi_j=\mu_j\rho'$, and observe that:
\[(v,c)\xtransrightarrow{\phi_0\cyclemark{\omega^{k_1}}\phi_1\cdots\phi_{i-1}\cyclemark{\omega^{k_i}}\mu_i}(v_i,c_i)
\xtransrightarrow{\rho\cyclemark{\omega^{k_{i+1}}}\phi_{i+1}\cdots\phi_{j-1}\cyclemark{\omega^{k_j}}\mu_j}(v_i,c_i)
\xtransrightarrow{\rho'\cyclemark{\omega^{k_{j+1}}}\phi_{j+1}\cdots\phi_{n-1}\cyclemark{\omega^{k_n}}\phi_n}(v',c').
\]
Thus the subpath $\rho\cyclemark{\omega^{k_{i+1}}}\phi_{i+1}\cdots\phi_{j-1}\cyclemark{\omega^{k_j}}\mu_j$ has
weight $0$ and rule \texttt{simplify} must apply:
\[
\chi\quad\leadsto\quad\phi_0\cyclemark{\omega^{k_1}}\phi_1\cdots\phi_{i-1}\cyclemark{\omega^{k_i}}\mu_i
\rho'\cyclemark{\omega^{k_{j+1}}}\phi_{j+1}\cdots\phi_{n-1}\cyclemark{\omega^{k_n}}\phi_n\]
which is a contradiction because we assumed that no rule can apply to $\chi$.

Consequently, for any $i\neq j$, we have $(v_i,c_i)\neq(v_j,c_j)$. But remember that $c_i\in B^+(v_i)$,
thus $(v_i,c_i)\in A$ where:
\[A=\bigcup_{v\in V} \{v\} \times B^+(v).\]
This shows that $n-1\leqslant|A|$. Indeed, by the pigeonhole principle, some pair $(v_i,c_i)$ would be repeated if $n-1>|A|$.
We can easily bound the size of $A$ using the bound on $B^+(v)$ from Lemma~\ref{lem:no_gather_obstruction}:
\[
|A|\leqslant\sum_{v\in V}|B^+(v)|\leqslant|V||\simpcycleposall|\sum_{v\in V}|\tau(v)|.
\]
Finally we have
\[
n\leqslant|V||\simpcycleposall|\sum_{v\in V}|\tau(v)|+1
    \leqslant|V||\simpcycleposall|\big(1+\sum_{v\in V}|\tau(v)|\Big)
\]
because $|V|\geqslant1$ and $|\simpcycleposall|\geqslant1$ unless there are no positive cycles,
in which case $n=0$ anyway.
\end{proof}

\begin{lem}[Length of equality-free computations]\label{lem:nf_eq_free_comp}
Let $\pi$ be a valid finite computation (without equality tests) from $(v,c)$ to $(v',c')$. Then there
  exists a folded path $\chi$ such that $\pathunfold{\chi}(c)$ is a valid computation from $(v,c)$ to $(v',c')$,
  the length of $\pathunfold{\chi}(c)$ is at most that of $\pi$
and the word length of $\chi$ is bounded by:
\[|V|+|V|^2|\simplecycleall|^2\Big(1+\sum_{v\in V}|\tau(v)|\Big)\]
\end{lem}

\begin{proof}
Let $\chi_0$ be the path defined by $\pi$: it is a word over alphabet $E$
and is thus a (trivial) folded path. By definition $\pathunfold{\chi_0(c)}=\pi$ is a valid
computation from $(v,c)$ to $(v',c')$ and the length of $\pathunfold{\chi_0(c)}$ is equal to that of $\pi$.
Let $\chi$ be any rewriting of $\chi_0$
such that no rule applies on $\chi$: it exists because there are no infinite
rewriting chains by Lemma~\ref{lem:rewrite_terminates}. By Lemma~\ref{lem:rewrite_sound},
$\pathunfold{\chi(c)}$ is still a valid computation from $(v,c)$ to $(v',c')$. Let $\omega$
be a simple cycle: note that it is either positive or negative, because rule \texttt{simplify}
removes zero-weight cycles. Then by Lemma~\ref{lem:no_rule_count_cycles},
the number of symbols of the form $\cyclemark{\omega^\cdot}$ appearing in $\chi$ is
bounded by\footnote{Since obviously $\max(|\simpcycleposall|,|\simpcyclenegall|)\leqslant|\simplecycleall|$.}:
\begin{equation}
|V||\simplecycleall|\Big(1+\sum_{v\in V}|\tau(v)|\Big)
\end{equation}
and thus the total number of symbols in $\chi$ of the form $\cyclemark{\omega^\cdot}$ for any $\omega$
is bounded by:
\begin{equation}\label{eq:bound_omegas}
|V| |\simplecycleall|^2 \Big(1+\sum_{v\in V}|\tau(v)|\Big).
\end{equation}
Furthermore, inbetween symbols of the form $\cyclemark{\omega^\cdot}$,
there can be subpaths consisting of symbols in $E$ only, so $\chi$ is of the form
\[\chi=\phi_0\cyclemark{\omega_1^{k_1}}\phi_1\cyclemark{\omega_2^{k_2}}\cdots\cyclemark{\omega_{n}^{k_{n}}}\phi_n\]
where $\phi_i\in E^*$ and $\omega_i\in\simplecycleall$ for all $i$. By the reasoning above, $n\leqslant\eqref{eq:bound_omegas}$.
Furthermore, by Lemma~\ref{lem:no_rule_small_E_words}, $\phi_i<|V|$ for all $i$.
It follows that the total length of $\chi$ is bounded by
\begin{align*}
(n+1)(|V|-1)+n&\leqslant|V|+n|V|\\
              &\leqslant|V|+|V|^2|\simplecycleall|^2\Big(1+\sum_{v\in V}|\tau(v)|\Big).
\end{align*}
Finally the length of $\pathunfold{\chi(c)}$ at most that of $\pi$
because the rewriting system does not increase the length of the path and the length of
$\pathunfold{\chi_0(c)}$ is equal to that of $\pi$.
\end{proof}

The main result of this section shows that any valid computation has an equivalent
valid computation given by a folded path whose length only depends on the automaton.

\begin{thm}[Length of computations]\label{th:nf_path}
Let $\pi$ be a valid finite computation from $(v,c)$ to $(v',c')$. Then there
exists a folded path $\chi$ such that $\chi(c)$ is a valid computation from $(v,c)$ to $(v',c')$,
the length of $\pathunfold{\chi(c)}$ is at most that of $\pi$
and the word length of $\chi$ is bounded by:
\[|E|\left(1+|V|+|V|^2|\simplecycleall|^2\Big(1+\sum_{v\in V}|\tau(v)|\Big)\right).\]
\end{thm}

\begin{proof}
Apply Lemma~\ref{lem:eq_test_isolate} to isolate the equality tests (at most $|E|$ of them)
and apply Lemma~\ref{lem:nf_eq_free_comp} to each equality-free subcomputation.
We can improve the bound slightly by noticing that there can only be up to $|E|$ equality-free subcomputations (and not $|E|+1$).
Indeed, if there are $|E|$ different equality tests in the path, there are no further edges available
for equality-free computations, and the word length is at most $|E|$.
\end{proof}

%%% Local Variables:
%%% mode: latex
%%% TeX-master: "Main-Revised.tex"
%%% End:

\section{Reachability with Parameterised Tests}
\label{sec:param-reach}
In this section we will show that both the reachability problem and
the generalised repeated control-state reachability problem for 1-CA with
parameterised tests are decidable, via a symbolic encoding of folded
paths, making use of the normal form from the previous section. The
result of this encoding is a formula of Presburger arithmetic.

Recall that
$C = \{ \cyclemark{\omega^k} : \omega \in \simplecycleall, k \in \N
\}$.
Let $C' = \{ \cyclemark{\omega^\cdot} : \omega \in
\simplecycleall\}$.
We define a \emph{path shape} to be a word over the alphabet
$E \cup C'$: $\xi = t_1\ldots t_n$ such that
$\pathend{t_i} = \pathstart{t_{i+1}}$, where
$\pathstart{\cyclemark{\omega^\cdot}} =
\pathend{\cyclemark{\omega^\cdot}} = \pathstart{\omega}$.
Given a path shape
$\xi = \gamma_0 \cyclemark{\omega_1^\cdot} \gamma_1 \ldots
\cyclemark{\omega_n^\cdot} \gamma_n$
with $\gamma_i \in E^*$, we write $\xi(k_1,\ldots,k_n)$ for the folded
path
$\gamma_0 \cyclemark{\omega_1^{k_1}} \gamma_1 \ldots
\cyclemark{\omega_n^{k_n}} \gamma_n$.
The advantage of working with path shapes rather than folded paths is that
the former are words over a finite
alphabet. 

\begin{lem}[Encoding computations]
  Given a 1-CA $\cC=(V,E,X,\lambda,\tau)$ with parameterised tests and
  configurations $(v,c)$ and $(v',c')$, and given a path shape
  $\xi = t_1 t_2 \ldots t_n \in (E \cup C')^*$, there exists a
  Presburger arithmetic formula
  $\varphi_{comp}^{(\xi),(v,c),(v',c')}(\vk,\vx)$, with free variables
  $\vx$ corresponding to the parameters $X$ and $\vk$ corresponding to
  exponents to be substituted in $\xi$, which evaluates to true if and
  only if $\pathunfold{\xi(\vk)}(c)$
  is a valid computation from $(v,c)$ to $(v',c')$.
\end{lem}

\begin{proof}
  Assume first that $\xi$ does not include any equality tests. We define a formula $\varphi_{valid,noeq}^{(t)}(\vk,\vx,y)$ which, given an equality-free symbol $t \in E \cup C'$ and an integer $y$, evaluates to true if and only if $\pathunfold{t(\vk)}(y)$ is a valid computation. There are two cases:
  \begin{itemize}
    \item $t \in E$. Then $\varphi_{valid,noeq}^{(t)}(\vx,y) \equiv y \geqslant 0 \wedge y + \pathweight{t} \geqslant 0 \wedge y \notin \tau(\pathstart{t})$.
    \item $t \in C'$, i.e., $t(\vk) = \cyclemark{\omega^k}$ for some simple cycle $\omega = e_1 e_2 \ldots e_\ell$ and $k \in \vk$. Then
      \begin{align*} \varphi_{valid,noeq}^{(t)}&(\vk,\vx,y) \equiv \forall k' \, (0 \leqslant k' < k) \Rightarrow \bigwedge_{i=1}^\ell \left ( y+ k' \pathweight{\omega} + \sum_{j=1}^{i-1} \pathweight{e_j} \geqslant 0 \wedge \right. \\
                                          &\left. y+ k' \pathweight{\omega} + \sum_{j=1}^{i-1} \pathweight{e_j} \notin \tau(\pathstart{e_i}) \right ) \wedge y + k \pathweight{\omega} \geqslant 0.
      \end{align*}
  \end{itemize}
  Note that for each edge $e \in E$, $\pathweight{e}$ is a constant,
  given by the automaton, and $\pathweight{\omega}$ is a shorthand for
  $\sum_{i=1}^\ell \pathweight{e_i}$, which is also a constant. So the
  only type of multiplication in the formula is by a constant.  A
  formula of the form $a \notin \tau(u)$ is a shorthand for
  $\bigwedge_{b \in \tau(u)} a \neq b$, which is clearly a Presburger
  arithmetic formula. Since $\cC$ has parameterised tests, in general
  some of these disequalities include variables from $\vx$. We can now
  define a formula with the required property in the case where $\xi$
  does not include any equality tests:
  \begin{align*}
    \varphi_{comp,noeq}^{(\xi),(v,c),(v',c')}&(\vk,\vx) \equiv \left ( \bigwedge_{i=1}^{n-1} \pathend{t_i} = \pathstart{t_{i+1}} \right ) \wedge
    \pathstart{t_1} = v \wedge \pathend{t_n} = v' \wedge \\
    & \sum_{i=1}^n \pathweight{t_i(\vk)} = c'-c \wedge \bigwedge_{i=1}^n \varphi_{valid,noeq}^{(t_i)}(\vk,\vx,c+\sum_{j=1}^{i-1} \pathweight{t_j(\vk)}),
  \end{align*}
  where we use the shorthand $\pathweight{s}$ for $s \in E \cup C$: if
  $s \in E$ then $\pathweight{s}$ is a constant as above, and if
  $s \in C$ then it is of the form $\cyclemark{\omega^k}$ and
  $\pathweight{s} = k \sum_{e \in \omega} \pathweight{e}$. Again, the
  only multiplications are by constants, so the resulting formula is a
  formula of Presburger arithmetic.

  Finally, in the case where $\xi$ includes equality tests, we split $\pathunfold{\xi}$ at the $t_i$ which are equality tests, and construct a formula $\varphi_{comp,noeq}$ as above for each equality-free part of $\xi$. $\varphi_{comp}^{(\xi),(v,c),(v',c')}(\vk,\vx)$ is the conjunction of these formulas.
\end{proof}

\begin{rem}[Removing the universal quantification]
    \label{rem:reachability_pure_existencial}
    For simplicity, we have used a universal quantifier in
    $\varphi_{valid,noeq}^{(t)}(\vk,\vx,y)$ to express that $k$ iterations of a cycle yield a valid computation.  In fact it is possible to
    rewrite $\varphi_{valid,noeq}^{(t)}(\vk,\vx,y)$ as a purely existential
    formula, with a polynomial blowup. Let $\omega=e_1\cdots e_\ell$ be a
    cycle and suppose we want to check that $\omega^k(y)$ is a valid computation. Let $u=\pathstart{e_i}$
    be a state on the cycle. First we need to express that the counter value at $u$ is never negative along $\omega^k(y)$. Since the counter value at $u$ is
    monotone during the $k$ iterations of the cycle (it increases if $\omega$ is
    positive and decreases if $\omega$ is negative), we only need
    check that it is nonnegative at the first and last iteration:
    \[y+\sum_{j=1}^{i-1}\pathweight{e_j}\geqslant0\wedge y+(k-1)\pathweight{\omega}+\sum_{j=1}^{i-1}\pathweight{e_j}\geqslant0.\]

    Next, for each $b\in\tau(u)$,
    we need to check that the cycle avoids $b$ in $u$. 
    Without loss of generality, assume that $\omega$ is positive. Then
    the counter value at $u$ increases after each iteration. 
    We can now perform a case analysis on the three ways to satisfy a disequality test during the $k$ iterations of $\omega$:
    \begin{itemize}
    \item The value at the first iteration is already bigger than $b$:
        \[y+\sum_{j=1}^{i-1}\pathweight{e_j}>b.\]
    \item The value at the last iteration is less than $b$:
      \[y+(k-1)\pathweight{\omega}+\sum_{j=1}^{i-1}\pathweight{e_j}<b.\]
    \item There is an iteration $k'$, with $0\leqslant k'<k-1$, at which the counter value is less than $b$,
        but where at the next iteration $k'+1$ the counter value is bigger than $b$:
        \begin{align*}
            \exists k' \, (0 \leqslant k' < k-1)
                &\wedge y+k'\pathweight{\omega}+\sum_{j=1}^{i-1}\pathweight{e_j}<b\\
                &\wedge y+(k'+1)\pathweight{\omega}+\sum_{j=1}^{i-1}\pathweight{e_j}>b.\\
        \end{align*}
    \end{itemize}

    Finally, we can use a conjunction over all states in $\omega$ to get a formula which is equivalent to $\varphi_{valid,noeq}^{(t)}(\vk,\vx,y)$ but has no universal quantifiers.
\end{rem}

\begin{lem}[Encoding reachability]
  \label{lem:reachability_encoding}
  Let $\cC = (V,E,X,\lambda,\tau)$ be a 1-CA with parameterised tests,
  and let $(v,c)$ and $(v',c')$ be given configurations of $\cC$. Then
  there exists a Presburger arithmetic formula
  $\varphi_{reach}^{(v,c),(v',c')}(\vx)$ which evaluates to true if
  and only if there is a valid computation from $(v,c)$ to $(v',c')$
  in $\cC$, as well as a formula $\varphi_{reach_+}^{(v,c),(v',c')}$
  which is true if and only if there is such a computation of length
  at least $1$.
\end{lem}

\begin{proof}
  Note that the bounds on the length of computations in 1-CA from the previous section do not depend on the values occurring in equality or disequality tests. That is, if there is a valid computation $(v,c) \xtransrightarrow{\text{ }\pi\text{ }} (v',c')$ for any given values of the parameters, then there is a folded path $\chi$ of word length at most $p(\cC)$ such that $(v,c) \xtransrightarrow{\pathunfold{\chi(c)}} (v',c')$ is a valid computation, where $p$ is the polynomial function given in Theorem~\ref{th:nf_path}. Equivalently, there is a path shape $\xi$ of word length at most $p(\cC)$ and there exist values $\vk$ such that $(v,c) \xtransrightarrow{\pathunfold{\xi(\vk)(c)}} (v',c')$ is a valid computation.

  Since path shapes are words over a finite alphabet, we can express this property as a finite disjunction
  \[ \varphi_{reach}^{(v,c),(v',c')}(\vx) \equiv \exists \vk \bigvee_{|\xi| \leqslant p(\cC)} \varphi_{comp}^{(\xi),(v,c),(v',c')}(\vk,\vx).  \]

  For $\varphi_{reach_+}$, we simply change the disjunction to be over all $\xi$ such that $1 \leqslant |\xi| \leqslant p(\cC)$.
\end{proof}

\begin{lem}[Encoding repeated control-state reachability]
  \label{lem:repeated_reachability_encoding}
  Let $\cC = (V,E,X,\lambda,\tau)$ be a 1-CA with parameterised tests,
  let $F \subseteq V$ be a set of final states, and let $(v,c)$ be the
  initial configuration of $\cC$. Then there exists a Presburger
  arithmetic formula $\varphi_{rep\text{-}reach}^{(v,c),(F)}(\vx)$ which
  evaluates to true if and only if there is a valid infinite
  computation $\pi$ which starts in $(v,c)$ and visits at least one
  state in $F$ infinitely often.
\end{lem}

\begin{proof}
  Suppose there is an infinite computation which starts in $(v,c)$ and visits a state $u \in F$ infinitely often. Equivalently, there is a counter value $d \in \N$ such that $(v,c) \longrightarrow^* (u,d)$ is a valid (finite) computation, and there is a cycle $\omega$ with $\pathstart{\omega} = u$ such that $\omega^k(d)$ is a valid computation for all $k \in \N$. There are two possible cases:
  \begin{itemize}
    \item $\pathweight{\omega} = 0$, so $\omega^k(d)$ is valid for all $k$ if and only if $\omega(d)$ is valid.
    \item $\pathweight{\omega} > 0$, so it might be possible to start
      from $(u,d)$ and follow the edges of $\omega$ a finite number of
      times before an obstruction occurs. However, if $\omega$ can be
      taken an arbitrary number of times, then the counter value will
      tend towards infinity, so we are free to choose $\omega$ to be
      an equality-free simple cycle, and $d$ to be high enough to
      guarantee that if $\omega$ can be taken once without
      obstructions, it can be taken infinitely many times.
  \end{itemize}
  The resulting formula is then
  \begin{align*}
    \varphi_{rep\text{-}reach}^{(v,c),(F)}(\vx) \equiv \, &\exists d \bigvee_{u \in F} \left ( \vphantom{\bigvee_{\omega \in \simpcycleposall}}
    \varphi_{reach}^{(v,c),(u,d)}(\vx) \wedge \left (\varphi_{reach_+}^{(u,d),(u,d)}(\vx) \vee  \right. \right. \\
    &\left. ( d>M(\vx) \wedge \exists d' \bigvee_{\omega \in \simpcycleposall}  \varphi_{comp,noeq}^{(\cyclemark{\omega^\cdot}),(u,d),(u,d')}(1,\vx)) \left.
    \vphantom{\varphi_{comp}^{(u,d')}} \right ) \right )
  \end{align*}
  where $M(\vx) = \max \left (\bigcup_{v \in V} \tau(v) \right ) - \sum \{ \pathweight{e} : e \in E, \pathweight{e} <0 \}$. The sum over negative edge weights ensures that the counter always stays above $\max \left (\bigcup_{v \in V} \tau(v) \right )$ along the computation $\omega(d)$, since each edge is taken at most once in $\omega$. Since $\omega$ is a positive cycle, this implies that the counter always stays above all bad values along $\omega(d^k)$ for each $k \in \N$, so no obstructions can occur.
\end{proof}

\begin{thm}[Decidability of reachability problems]
  Both the reachability problem and the generalised repeated control-state reachability problem are decidable for 1-CA with parameterised tests.
\end{thm}

\begin{proof}
  Given a 1-CA $\cC = (V,E,X,\lambda,\tau)$ with parameterised tests and configurations $(v,c)$ and $(v',c')$, to check if there exist values for the parameters $X$ such that there is a valid computation from $(v,c)$ to $(v',c')$, we use Lemma~\ref{lem:reachability_encoding} to construct the formula $\exists \vx \, \varphi_{reach}^{(v,c),(v',c')}(\vx)$.

  To solve the generalised repeated control-state reachability problem for a 1-CA $\cC = (V,E,X,\lambda,\tau)$ with sets of final states $F_1,\ldots,F_n \subseteq V$ and initial configuration $(v,c)$, note that this problem can easily be reduced to the simpler case where $n=1$, using a translation similar to the standard translation from generalised B\"uchi automata to B\"uchi automata. In the case where $n=1$, we can use Lemma~\ref{lem:repeated_reachability_encoding} to construct the formula $\exists \vx \, \varphi_{rep\text{-}reach}^{(v,c),(F_1)} (\vx)$.
\end{proof}

\begin{cor}[Decidability of model checking flat Freeze LTL]
  The existential model checking problem for flat Freeze LTL on 1-CA is decidable.
\end{cor}

%%% Local Variables:
%%% mode: latex
%%% TeX-master: "Main-Revised.tex"
%%% End:

\section{Conclusion}

The main result of this paper is that the model checking problem for
the flat fragment of Freeze LTL on one-counter automata is decidable.
We have concentrated on showing decidability rather than achieving
optimal complexity.  For example, we have reduced the model checking
problem to the decision problem for the class of sentences of
Presburger arithmetic with quantifier prefix $\exists^* \forall^*$.
We explained in Remark~\ref{rem:reachability_pure_existencial}
that in fact the reduction can be refined to yield a (polynomially larger)
purely existential sentence.

Another important determinant of the complexity of our procedure is
the dependence of the symbolic encoding of computations (via path
shapes) in Section~\ref{sec:param-reach} on the number of simple
cycles in the underlying control graph of the one-counter
automaton. The number of such cycles may be exponential in the number
of states. It remains to be seen whether it is possible to give a
more compact symbolic representation, e.g., in terms of the Parikh
image of paths. As it stands, our procedure for model checking flat Freeze LTL formulas on classical one-counter automata works as follows. From the
flat Freeze LTL formula and the automaton, we build a one-counter automaton with parameterised tests 
(of exponential size). We then guess the normal form of the
path shapes (of exponential size in the size the automaton). We
finally check the resulting existential Presburger formula.  Since the
Presburger formula has size double exponential in the size of the
input, we get a naive upper bound of 2NEXPTIME for
our algorithm.  Improving this bound is a subject of ongoing work.

Another interesting complexity question concerns configuration
reachability in one-counter automata with non-parameterised equality
and disequality tests.  For automata with only equality tests and with
counter updates in binary, reachability is known to be
NP-complete~\cite{HKOW}.  If inequality tests are allowed then
reachability is PSPACE-complete~\cite{FearnleyJ15}.  Now automata with
equality and disequality tests are intermediate in expressiveness
between these two models and the complexity of reachability in this
case is open as far as we know.

%%% Local Variables:
%%% mode: latex
%%% TeX-master: "Main-Revised.tex"
%%% End:

\bibliographystyle{plain}
\bibliography{Bibliography.bib}

\end{document}